\documentclass[reprint,amsfonts, amssymb, amsmath, prx,nofootinbib, superscriptaddress, twocolumn,longbibliography,aps]{revtex4-2}
\usepackage{float}
\makeatletter
\let\newfloat\newfloat@ltx
\makeatother

\usepackage[utf8]{inputenc}
\usepackage[english]{babel}
\usepackage{graphics}
\usepackage{selinput}
\usepackage[shortlabels]{enumitem}

\usepackage{braket}
\usepackage{amsthm}
\usepackage{amssymb}
\DeclareMathOperator{\Tr}{Tr}
\usepackage{amsmath}
\usepackage{mathtools}
\usepackage[caption=false,font=small]{subfig}
\usepackage{physics}
\usepackage{booktabs}

\usepackage{pifont}
\usepackage{xcolor}
\usepackage{graphicx}
\usepackage{adjustbox}
\usepackage{placeins}
\usepackage[T1]{fontenc}
\usepackage{csquotes}
\usepackage{bm}
\usepackage{bbm}
\usepackage{nameref}

\usepackage[makeroom]{cancel}
\usepackage[toc,page]{appendix}
\usepackage[colorlinks=true,allcolors=blue]{hyperref}

\usepackage{tikz}
\usetikzlibrary{
    positioning,
    fit,
    backgrounds,
    decorations.pathmorphing,
    arrows.meta,
    calc
}

\newtheorem{theorem}{Theorem}

\newtheorem*{criterion}{Criterion}

\newtheorem{proposition}{Proposition}

\newtheorem{definition}{Definition}

\newcounter{oldeqsave}\newcounter{olddefsave}

\makeatletter
\newif\ifeqcontrib@this
\newif\ifeqcontrib@any
\eqcontrib@thisfalse
\eqcontrib@anyfalse

\begin{document}
\preprint{APS/123-QED}

\title{
Planckian bounds on thermalization times of strongly coupled open quantum systems
}

\author{Jesús Jiménez-Rodríguez}
\email{jesus.jimenez.rodriguez@uab.cat}
\affiliation{Física Teòrica: Informació i Fenòmens Quàntics, Departament de Física, Universitat Autònoma de Barcelona, 08193 Bellaterra (Barcelona), Spain}

\author{John Calsamiglia}
\email{john.calsamiglia@uab.cat}
\affiliation{Física Teòrica: Informació i Fenòmens Quàntics, Departament de Física, Universitat Autònoma de Barcelona, 08193 Bellaterra (Barcelona), Spain}

\author{Martí Perarnau-Llobet}
\email{marti.perarnau@uab.cat}
\affiliation{Física Teòrica: Informació i Fenòmens Quàntics, Departament de Física, Universitat Autònoma de Barcelona, 08193 Bellaterra (Barcelona), Spain}

\date{\today}

\begin{abstract}
We consider whether the thermalization of strongly coupled open quantum systems, whose equilibrium state is the mean-force Gibbs (MFG) state, is constrained by a fundamental Planckian time scale, $\hbar/k_BT$. By requiring that the same process thermalizes a family of nearby system Hamiltonians, we obtain a lower bound on the thermalization time, irrespective of the specific system–bath dynamics. The bound depends only on the distinguishability of the corresponding MFG states, and can therefore be computed from equilibrium properties alone. In its simplest form, it requires only a static susceptibility and an equilibrium variance of a system observable. We characterize this bound from weak to strong coupling in exactly solvable fermionic models, in the spin--boson model,
and in a spin coupled to a two-dimensional Heisenberg antiferromagnet. These models show that strong coupling can lift the Planckian constraint when the MFG state retains little information about the system Hamiltonian, while preserving it when that information is carried by system observables with substantial equilibrium fluctuations at frequencies below $k_BT/\hbar$. The bound is complementary to open-system speed limits, since it constrains the acquisition of information rather than the preparation of the target, and can remain informative at strong coupling even when the latter become trivial.
\end{abstract}
\maketitle

\section{Introduction}

How fast can a quantum system thermalize? Understanding how many-body systems equilibrate and thermalize is a central problem in nonequilibrium physics~\cite{Eisert2015QuantumManyBodyOutOfEquilibrium,GogolinEisert2016}. Typicality and the eigenstate thermalization hypothesis explain why local observables thermalize~\cite{Deutsch1991,Srednicki1994,Rigol2008,Linden2009,Short2012QuantumEquilibration,ReimannKastner2012}, but much less is known about how long this takes. General bounds on equilibration times tend to be either unrealistically long or  short~\cite{Goldstein2013TimeScales,Malabarba2014,Goldstein2015,Reimann2016}, and estimates for physically relevant observables have only recently become available~\cite{GarciaPintos2017Equilibration,deOliveira2018Equilibration,Wilming2018EquilibrationTimes,Alhambra2020TimeEvolution,Wang2025Estimate}, with the underlying assumptions and system-size dependence still debated~\cite{Heveling2020Compelling,Heveling2020Comment,Schiulaz2019Thouless} (see Ref.~\cite{Gong2022Bounds} for a review). In open systems~\cite{petruccione}, relaxation can be slowed down by long-lived collective modes and strong decoherence~\cite{Minganti2018,Wei2023MultipleSteadyStates,Windt2024FermionicMatterWave,PhysRevA.106.042209,hierarchical}, or anomalously accelerated, as in the Mpemba effect~\cite{LuRaz2017,Kumar2020,Carollo2021Mpemba,Ares2023Asymmetry}. These time scales are probed in platforms where relaxation is engineered or monitored~\cite{Trotzky2012Relaxation,Kaufman2016Thermalization,Pekola2015QuantumThermodynamics,Poyatos1996ReservoirEngineering,Barreiro2011OpenSystemSimulator}, and controlling them is relevant for quantum technologies, from dissipative state preparation~\cite{Diehl2008,Verstraete2009} to qubit reset~\cite{Aamir2025}.

A central question, both fundamentally and practically, is therefore whether system--environment engineering can accelerate relaxation indefinitely, or whether quantum mechanics imposes a minimum thermalization time. At finite temperature, a natural candidate is the Planckian time $\tau_{\rm Pl}=\hbar\beta$, with $\beta=1/(k_BT)$, first identified in the transport properties of strongly correlated metals \cite{Zaanen2004WhyTemperatureHigh,bruin2013similarity,hartnoll2015theory,Fratini2026,Fratini2026StrangeMetal}. This universal scale is conjectured to set a bound on the speed of  dissipation and thermalization in quantum many-body systems~\cite{Hartnoll2022PlanckianDissipation,Lucas2019OperatorSize,mousatov2020planckian,delacretaz2025bound,Abiuso2025PlanckianBound}, chaos~\cite{Maldacena2016,Tsuji2018Bound,Murthy2019Bounds,Pappalardi2022QuantumBounds,Delacretaz2022}, and information erasure~\cite{Rolandi2023finitetimelandauer}.
Here we ask whether this scale constrains the thermalization of systems that are strongly coupled to their environment.

Standard tools to constrain dynamical time scales are quantum speed limits~\cite{Mandelstam1945TimeEnergy,Margolus1998MaximumSpeed,DeffnerCampbell2017,Gong2022Bounds}, which have been extended to open systems~\cite{Taddei2013QuantumSpeedLimit,delCampo2013,Deffner2013QuantumSpeedLimit,Funo2019SpeedLimit,Shiraishi2021SpeedLimit,Das2021ThermodynamicsQSL,nishiyama2026unifiedspeedlimitsclassical} and are complemented in many-body systems via Lieb--Robinson bounds~\cite{Lieb1972FiniteGroupVelocity,Nachtergaele2006LiebRobinson}.
For open systems,  speed limits typically compare the distance traveled  by the reduced state of the system with  an average dynamical rate. The latter is controlled by energy variances, interaction norms, or energy currents~\cite{nishiyama2026unifiedspeedlimitsclassical,Deffner2013QuantumSpeedLimit,Shiraishi2021SpeedLimit,Das2021ThermodynamicsQSL}.
These quantities can grow with the coupling strength, the reservoir size, or the bandwidth of the bath~\cite{Shiraishi2021SpeedLimit}, even when the reduced dynamics does not depend on them.
Speed limit bounds can then become vanishingly small precisely in the strongly coupled, macroscopic regime, where relaxation is not necessarily fast~\cite{Minganti2018,Wei2023MultipleSteadyStates,Windt2024FermionicMatterWave,PhysRevA.106.042209,hierarchical}. This motivates constraints that remain informative without requiring a bound on the bath properties.

There is also a conceptual reason why speed limits are not the natural tool for thermalization. They bound the cost of preparing one given target from one given initial state. But if the system Hamiltonian $H_S$ is known in advance, a device could store copies of its Gibbs state $\omega(\beta,H_S)=e^{-\beta H_S}/\Tr[e^{-\beta H_S}]$ and swap them with the system arbitrarily quickly, provided the interaction strength is unrestricted~\cite{collisional}. Such a protocol would prepare the correct state for that particular Hamiltonian, yet it would not constitute genuine thermalization: it fails as soon as $H_S$ is slightly changed.
Thermalization is better understood as a robust process, in which the same mechanism (the same bath, interaction, and initial state) brings the system close to the appropriate equilibrium state whichever Hamiltonian in a family of nearby ones governs it.

Following Ref. \cite{Abiuso2025PlanckianBound},  this robustness requirement turns thermalization into an information-processing task. If two Hamiltonians have distinguishable equilibrium states, any process that thermalizes both must produce distinguishable final states; that is, it must encode which Hamiltonian is present. Quantum estimation theory limits how fast this information can be acquired~\cite{Paris2009QuantumEstimation,Boixo2007GeneralizedLimits}, which yields a lower bound on the thermalization time. This approach  leads to a rigorous Planckian bound $\tau\geq\tau_{\rm Pl}/2$ for Gibbs state preparation~\cite{Abiuso2025PlanckianBound}.
The same logic, namely lower-bounding a dynamical time scale by the sensitivity of the steady state to a Hamiltonian parameter through quantum estimation theory, has recently been applied to the relaxation and photon-correlation times of Markovian driven-dissipative systems with non-thermal steady states~\cite{Grecki2026Time}.

In this work, we extend this approach to the  mean-force Gibbs (MFG) state,  that is,  the reduced state of the global system--bath Gibbs state, which describes thermalization at arbitrary system--environment coupling~\cite{Miller2018,trushechkin2022open,pwws-21cx}.
The core question becomes whether the MFG state retains enough information about $H_S$ to impose a nontrivial bound, and whether that bound remains Planckian. The MFG state and its dependence on $H_S$ are hard to compute, and are known in closed form only in limiting regimes or solvable models~\cite{CresserAnders2021MFG,Chiu2022MFGNumerics}. We therefore avoid the MFG state itself and express the bound through static susceptibilities of system observables, i.e., the response of their equilibrium expectation values to changes of $H_S$, together with equilibrium variances. These are standard quantities of linear-response theory~\cite{KuboTodaHashitsume1991}, and they are equilibrium correlations of the system--bath compound.

This approach enables us to show when the bound is Planckian. The susceptibility is an average of the equilibrium autocorrelation
of the perturbing observable over a window of width $\hbar/k_BT$. The bound then becomes Planckian when some perturbation of $H_S$ fluctuates appreciably in equilibrium and its fluctuations remain correlated over this window, and it falls below the Planckian scale when the relevant fluctuations are small or decorrelate faster than $\hbar/k_BT$.
We test this criterion in three settings: (i)~in exactly solvable fermionic models, where the bound can be compared with exact thermalization times and with speed limits, (ii)~in the spin--boson model, whose MFG state we compute with numerically exact methods at arbitrary coupling~\cite{Chiu2022MFGNumerics,Strathearn2018TEMPO}, and (iii)~in a two-dimensional Heisenberg antiferromagnet, whose real-time dynamics is out of reach, and where we evaluate the bound via  quantum Monte Carlo methods.
In general, the bound need not be tight, since it constrains the acquisition of equilibrium information rather than the full relaxation dynamics, but it remains informative at arbitrary coupling.  Furthermore, it can remain finite where speed limits based on the interaction energy vanish, making the two approaches  complementary.

The rest of this article is organized as follows. Section~\ref{sec:frame} introduces the strong-coupling setting, the information-theoretic notion of thermalization, and the resulting Planckian bound, and compares it with open-system speed limits. Section~\ref{sec:mfg_sensitivity} expresses the sensitivity of MFG states in terms of equilibrium susceptibilities and identifies the most demanding Hamiltonian perturbations. Section~\ref{sec:fermionic} applies the bound to the resonant-level model and to a double quantum dot, comparing it with exact thermalization times and with open-system speed limits. Section~\ref{sec:spin_boson} applies it to the spin--boson model from weak to ultrastrong coupling. Section~\ref{sec:heisenberg} evaluates it from equilibrium data alone for a spin in a two-dimensional antiferromagnet, whose dynamics is out of reach. Section~\ref{sec:conclusion} summarizes our results. Proofs and details of the fermionic models are given in Appendices~\ref{app:mfg_sensitivity} and \ref{app:fermionic}.

\section{Thermalization bounds beyond weak coupling}\label{sec:frame}

We connect thermalization times to equilibrium-state distinguishability by requiring a common mechanism to prepare the appropriate targets for a family of nearby Hamiltonians.

\subsection{Open quantum systems at strong coupling}
\label{subsec:strong_coupling}

We consider a system $S$ coupled to an environment $B$, with total Hamiltonian
\begin{equation}
    H_{\rm tot}
    =
    H_S+H_B+V,
    \label{eq:total_hamiltonian}
\end{equation}
where $H_S$ is the system Hamiltonian, $H_B$ is the bath Hamiltonian, and $V$ is the system--bath interaction. We write $V=gV_0$, where $V_0$ is a fixed interaction operator and $g\geq0$ controls the system--bath coupling strength. The compound $S+B$ evolves unitarily according to
\begin{equation}
    \dot{\rho}_{SB}(t)
    =
    -\frac{i}{\hbar}
    \left[
        H_{\rm tot},
        \rho_{SB}(t)
    \right],
    \quad
    \rho_S(t)
    =
    \Tr_B[
        \rho_{SB}(t)
    ] .
    \label{eq:oqs_reduced_dynamics}
\end{equation}
Keeping the joint dynamics explicit allows us to treat strong coupling without assuming Markovianity or a master equation for the reduced state. We impose no bound on the interaction strength or the size of the environment.

At inverse temperature $\beta$, we define the thermalization target as the system marginal of the global Gibbs state of the interacting compound,
\begin{equation}
\omega(\beta,H_{\rm tot})
=
\frac{
e^{-\beta H_{\rm tot}}
}{
\Tr[
e^{-\beta H_{\rm tot}}
]
},
\label{eq:global_gibbs_state}
\end{equation}
namely,
\begin{equation}
\tilde{\omega}(\beta,H_S)
=
\Tr_B
\left[
\omega(\beta,H_{\rm tot})
\right],
\label{eq:reduced_global_gibbs_state}
\end{equation}
where the dependence on $H_S$ is understood at a fixed $H_B$ and $V$. This reduced equilibrium state is the mean-force Gibbs (MFG) state \cite{Miller2018,trushechkin2022open}. It coincides with the bare Gibbs state $\omega(\beta,H_S)$ when the interaction with the bath is negligible ($g\to0$).  At finite coupling, the interaction shapes the equilibrium properties of $S$, and the resulting state can differ substantially from the bare Gibbs state.

Choosing $\tilde{\omega}(\beta,H_S)$ as the target does not assume that arbitrary system--bath dynamics converges to it. Our bounds apply to processes that approach this target with the accuracy and Hamiltonian robustness specified below.

\subsection{An information-theoretic notion of thermalization}
\label{subsec:it_thermalization}

Following Ref.~\cite{Abiuso2025PlanckianBound}, we require a common physical mechanism to approach the corresponding equilibrium states for a nontrivial family of nearby system Hamiltonians. The bath Hamiltonian $H_B$, interaction $V$, and initial joint state $\rho_{SB}(0)$ are held fixed across this family: only $H_S$ changes. This requirement prevents the mechanism from encoding the choice of target state in its preparation and makes thermalization a task of acquiring information about $H_S$.

\begin{definition}[Robust thermalization]
\label{def:thermalization}
Fix the bath Hamiltonian $H_B$, the interaction $V$, and the initial state $\rho_{SB}(0)$, and let $\rho_S(t,H_S)$ be the reduced state obtained from Eq.~\eqref{eq:oqs_reduced_dynamics} with system Hamiltonian $H_S$. The process thermalizes robustly at time $\tau$, with accuracy $\varepsilon>0$ over the Hamiltonian range
\begin{equation}
\mathcal{B}_\delta(\bar H_S)\coloneq\{H_S:\ \|H_S-\bar H_S\|\le\delta\},
\label{eq:smooth_thermalization}
\end{equation}
if
\begin{equation}
D\big(\rho_S(\tau,H_S),\tilde\omega(\beta,H_S)\big)\le\varepsilon
\qquad \forall\, H_S\in\mathcal{B}_\delta(\bar H_S).
\label{eq:approx_thermalization}
\end{equation}
Here $D(\rho,\sigma) \coloneq \arccos(\sqrt{F(\rho,\sigma)})$ is the Bures angle, with $F(\rho, \sigma)=\Tr(\sqrt{\sqrt{\sigma}\rho\sqrt{\sigma}})^2$ the quantum fidelity, and $\|A\|:=\lambda_{\rm max}(A)-\lambda_{\rm min}(A)$ is the spectral seminorm, with $\lambda_i(A)$ the eigenvalues of $A$ \footnote{This seminorm is invariant under $A\to A+c\,\mathbb{I}$, which changes neither the equilibrium state nor the dynamics, and it equals $2\min_c\|A+c\,\mathbb{I}\|_{\rm op}$. It is the natural distance between Hamiltonians here, and the one for which the generalized Heisenberg limit takes the form used in Sec.~\ref{subsec:locally_exact}}.
\end{definition}
We define the thermalization time as the first such time,
\begin{align}
\tau_{\delta,\varepsilon}\coloneq\min\big\{t\ge0:\ &D\big(\rho_S(t,H_S),\tilde\omega(\beta,H_S)\big)\le\varepsilon\nonumber\\
&\forall\,H_S\in\mathcal{B}_\delta(\bar H_S)\big\},
\label{eq:thermalization_time}
\end{align}
with $\tau_{\delta,\varepsilon}=\infty$ if no such time exists. This definition requires neither convergence nor the absence of recurrences. Requiring instead that the state remain within $\varepsilon$ of its target at all later times would need such assumptions and would define a time no smaller than $\tau_{\delta,\varepsilon}$, so the bound below applies to it as well.

Definition~\ref{def:thermalization} combines finite accuracy with robustness to changes in the system Hamiltonian. If $\delta=0$, the problem reduces to preparing a single known target state.
If $\varepsilon$ is comparable to or larger than the Bures angles between the equilibrium targets in $\mathcal{B}_\delta(\bar H_S)$, the process need not resolve which Hamiltonian is present. Otherwise, thermalization requires the dynamics to acquire information about the system Hamiltonian.

The extension of Ref.~\cite{Abiuso2025PlanckianBound} to strong coupling enters through the target: the bare Gibbs state is replaced by the MFG state in Eq.~\eqref{eq:reduced_global_gibbs_state}. When two equilibrium targets remain distinguishable at the prescribed accuracy, thermalization must produce distinguishable reduced states through the joint evolution of $S+B$. The quantum limit on how quickly this evolution can encode the difference between the Hamiltonians yields the time bound below, proved in Ref.~\cite{Abiuso2025PlanckianBound}.

\subsection{Planckian thermalization bound}
\label{sec:planckian_bound}

\begin{theorem}[Thermalization bound at strong coupling] \label{thm:planckian_bound}
Let $\tau_{\rm Pl}=\hbar\beta$ be the Planckian time. For any process satisfying Definition~\ref{def:thermalization}, the thermalization time of Eq.~\eqref{eq:thermalization_time} obeys
\begin{align}
    & {\tau_{\delta,\varepsilon}}
    \geq
    \tau_{\rm Pl}\,
    \chi(\bar H_S,\delta,\varepsilon),  \quad \quad \text{with \ }
    \label{eq:chi_def}
    \\
    & \chi(\bar H_S,\delta,\varepsilon)
    \coloneq
    \nonumber\max_{\substack{H_S^{(1,2)}\in\mathcal{B}_\delta\\ {H_S^{(1)}\neq H_S^{(2)}}}}
    \left[
    \frac{
        2D\!\left(
            \tilde\omega^{(1)},\tilde\omega^{(2)}
        \right)
        -4\varepsilon
    }{
        \beta\,
        \|H_S^{(1)}-H_S^{(2)}\|
    }
    \right]_{+},
    \nonumber
\end{align}
where $\tilde\omega^{(i)}\equiv\tilde\omega(\beta,H_S^{(i)})$ and $[x]_+\coloneq\max\{x,0\}$.
\end{theorem}

The dimensionless factor $\chi(\bar H_S,\delta,\varepsilon)$ quantifies the equilibrium sensitivity relevant to thermalization over~$\mathcal{B}_\delta$. It
is positive whenever the equilibrium targets of some pair of Hamiltonians in $\mathcal{B}_\delta(\bar H_S)$ are separated by more than $2\varepsilon$ in Bures angle, so that the task requires more than preparing a single state. The factor satisfies $\chi(\bar H_S,\delta,\varepsilon)\leq {\big[}\frac{1}{2}-\frac{2\varepsilon}{\beta\delta}{\big]_+}$, because the Bures angle between two equilibrium targets is at most $\beta/4$ times the distance between their Hamiltonians (last inequality of Eq.~\eqref{eq:qfi_hierarchy}), and this distance is at most $2\delta$. Hence, a nontrivial lower bound requires $\varepsilon< \beta\delta/4$, which makes explicit the relation between the required accuracy and the Hamiltonian range.

At strong coupling, the question is therefore whether the MFG states retain enough information about $H_S$ to remain distinguishable at the prescribed accuracy, and whether the resulting bound remains of the order of $\tau_{\rm Pl}$.
We next examine the bound in two physically relevant regimes: locally exact thermalization and approximate thermalization at finite error.

\subsubsection{Locally exact thermalization and quantum Fisher information}
\label{subsec:locally_exact}

Let us now consider locally exact thermalization, obtained by shrinking the ball of Hamiltonians in Definition~\ref{def:thermalization} to an infinitesimal neighborhood of $\bar H_S$, $\delta\to0$, while keeping the error negligible, $\varepsilon\ll\beta\delta$. In this limit, the reduced state must reproduce the target equilibrium state to first order in $H_S-\bar H_S$. We parametrize the perturbations as
\begin{equation}
    H_S(\delta,\kappa)=\bar H_S+\delta\kappa,
    \qquad  \|\kappa\|=1,
    \label{eq:local_hamiltonian_perturbation}
\end{equation}
where $\kappa$ is a Hermitian operator and $\delta$ is infinitesimal. Fixing $\|\kappa\|=1$ entails no loss of generality, since a rescaling of $\kappa$ can be absorbed into $\delta$. Then $\|H_S-\bar H_S\|=|\delta|$, so that $\delta$ sets the size of the perturbation and $\kappa$ its direction. We use this normalization throughout. The bound in Eq.~\eqref{eq:chi_def} then becomes $\tau\geq\tau_{\rm Pl}\widetilde\chi(\bar H_S)$, with
\begin{align}
    \widetilde\chi(\bar H_S)
    &\coloneq
    \lim_{{\substack{\delta\to0\\ \varepsilon/(\beta\delta)\to0}}}
    \chi(\bar H_S,\delta,\varepsilon)
    =\max_{{\|\kappa\|=1}}
    \frac{\sqrt{\mathcal F^{\rm th}_\kappa(\beta,\bar H_S)}}{\beta}.
    \label{eq:local_chi}
\end{align}
Here $\mathcal F^{\rm th}_\kappa(\beta,\bar H_S)$ is the quantum Fisher information (QFI) of the MFG family $\tilde{\omega}(\beta,H_S(\delta,\kappa))$ with respect to $\delta$, evaluated at $\delta=0$~\cite{Paris2009QuantumEstimation}. We used the fact that the QFI of a parametric state $\rho_\delta$ is related to its local sensitivity to changes in $\delta$, measured by the Bures angle, through $\mathcal F[\rho_\delta]|_{\delta=0}=4\lim_{\delta\to0}[D(\rho_0,\rho_\delta)/\delta]^2$~\cite{Scandi2025QFIDynamicalNature}.

This result has a direct metrological interpretation. Locally exact thermalization requires the dynamical QFI $\mathcal F^{\rm dyn}_\kappa(\tau,\bar H_S)$, defined analogously for the output family $\rho_S(\tau,H_S(\delta,\kappa))$, to coincide with $\mathcal F^{\rm th}_\kappa(\beta,\bar H_S)$. The generalized Heisenberg limit, however, imposes $\mathcal F^{\rm dyn}_\kappa(\tau,\bar H_S)\leq\tau^2\|\kappa\|^2/\hbar^2=\tau^2/\hbar^2$~\cite{Boixo2007GeneralizedLimits}. The dynamics must therefore have enough time to acquire the Hamiltonian information present in equilibrium; maximizing over $\kappa$ yields Eq.~\eqref{eq:local_chi}. Ref.~\cite{Abiuso2025PlanckianBound} already noted that this locally exact argument holds for any target $\tilde\omega(H_S)$, including reduced Gibbs states; here $\tilde\omega$ is the MFG state, and Sec.~\ref{sec:mfg_sensitivity} shows how to evaluate its QFI. In the models below, a nonzero QFI of the MFG state thus provides a nontrivial lower bound on thermalization time at strong coupling, without solving the transient dynamics.

\subsubsection{Approximate thermalization}
\label{subsec:approximate_thermalization}
The limit of locally exact thermalization yields simple bounds and an intuitive interpretation in terms of QFI. The more general bound in Eq.~\eqref{eq:chi_def} follows from the same geometrical argument for finite variations of $H_S$, while allowing a finite error $\varepsilon>0$ in reaching the target equilibrium state. This is essential in applications, since thermalization in finite time is generally approximate rather than exact. We use this finite-error form to compare thermalization times at a prescribed accuracy with their lower bounds in the physical models considered below.

\subsection{Comparison with open-system speed limits}
\label{subsec:speed_limits_general}

Quantum speed limits provide another lower bound on the thermalization time~\cite{Mandelstam1945TimeEnergy,Margolus1998MaximumSpeed,Taddei2013QuantumSpeedLimit,Deffner2013QuantumSpeedLimit}. They bound how fast a state can change, and thermalization requires the reduced state to change from $\rho_S(0)$ to within $\varepsilon$ of its target $\tilde\omega(\beta,H_S)$. By the triangle inequality, the reduced state must then travel at least the Bures angle
\begin{equation*}
    \mathcal D(H_S)\coloneq\big[D\big(\rho_S(0),\tilde\omega(\beta,H_S)\big)-\varepsilon\big]_+
\end{equation*}
in the time $\tau_{\delta,\varepsilon}$. Its speed is at most $\sqrt{\mathcal F^S(t)}/2$, where $\mathcal F^S(t)$ is the QFI of $\rho_S(t)$ with respect to time~\cite{Taddei2013QuantumSpeedLimit,nishiyama2026unifiedspeedlimitsclassical}. Since $H_B$ alone does not change the reduced state, $\hbar\sqrt{\mathcal F^S(t)}/2$ is in turn at most the standard deviation $\delta E_S(t)$ of $H_S+V$ in the global state~\cite{Shiraishi2021SpeedLimit}, and $\delta E_S(t)\leq\|H_S+V\|/2$. Integrating the speed over $[0,\tau_{\delta,\varepsilon}]$ gives, for every $H_S\in\mathcal B_\delta(\bar H_S)$,
\begin{equation}
    \frac{2\hbar\,\mathcal D(H_S)}{\|H_S+V\|}
    \leq\frac{\hbar\,\mathcal D(H_S)}{\langle\delta E_S\rangle_{\tau_{\delta,\varepsilon}}}
    \leq\frac{2\,\mathcal D(H_S)}{\big\langle\sqrt{\mathcal F^S}\big\rangle_{\tau_{\delta,\varepsilon}}}
    \leq\tau_{\delta,\varepsilon},
    \label{eq:ss_chain}
\end{equation}
where $\langle\cdot\rangle_{\tau_{\delta,\varepsilon}}$ denotes the time average over $[0,\tau_{\delta,\varepsilon}]$ along the exact trajectory for $H_S$. Maximizing the middle form over the ball gives the Shiraishi--Saito (SS) bound~\cite{Shiraishi2021SpeedLimit}
\begin{equation}
    \tau_{\delta,\varepsilon}\geq\tau_{\rm SS}\coloneq\max_{H_S\in\mathcal B_\delta(\bar H_S)}\frac{\hbar\,\mathcal D(H_S)}{\langle\delta E_S\rangle_{\tau_{\delta,\varepsilon}}}.
    \label{eq:ss_bound}
\end{equation}
The three forms in Eq.~\eqref{eq:ss_chain} use increasing knowledge of the process. The first needs only the operator $H_S+V$, the second the energy fluctuations along the global trajectory, and the third the full reduced trajectory, which already determines $\tau_{\delta,\varepsilon}$. All three depend on the preparation $\rho_S(0)$ and on the generator $H_S+V$.

These bounds can lose their force at strong coupling. Writing $V=gV_0$, we have $\|H_S+V\|\geq g\|V_0\|-\|H_S\|$. The static form therefore vanishes as $1/g$ independently of the dynamics, and $\tau_{\rm SS}$ does too whenever the fluctuations of $V_0$ along the trajectory do not vanish. This is the lack of robustness anticipated in the Introduction. Speed limits can vanish even when the thermalization time grows with the coupling, as in the DQD (Sec.~\ref{subsec:speed_limits}).

Theorem~\ref{thm:planckian_bound} also bounds a rate of change of Bures angle, but between a different pair of states. Instead of following one reduced state from $\rho_S(0)$ to its target, it compares the evolutions under two Hamiltonians of the ball, which must end near distinguishable targets. The two total Hamiltonians differ only by $H_S^{(1)}-H_S^{(2)}$, so the rate at which they become distinguishable is set by $\|H_S^{(1)}-H_S^{(2)}\|$ alone, independently of $\bar H_S$, $V$, the bath, and $\rho_S(0)$. A speed limit is thus a \emph{preparation} cost. Theorem~\ref{thm:planckian_bound} is an \emph{information} cost, the time needed for the reduced state to encode which Hamiltonian is present.

The two costs behave differently. Our bound vanishes for a single Hamiltonian, $\delta=0$, where speed limits remain nonzero. In the locally exact limit of Sec.~\ref{subsec:locally_exact}, $\delta\to0$ with $\varepsilon/(\beta\delta)\to0$, it can instead remain finite. Conversely, it depends on the interaction only through the targets, and it vanishes at strong coupling only if these become indistinguishable at accuracy $\varepsilon$, as in the RLM (Sec.~\ref{subsec:rlm}). Neither bound implies the other. Both bound $\tau_{\delta,\varepsilon}$, and the larger one applies. Table~\ref{tab:bound_nature} summarizes these differences, and Sec.~\ref{subsec:speed_limits} compares the two bounds in the fermionic models.

\begin{table}[t]
    \caption{\label{tab:bound_nature}%
    Speed limits of Eq.~\eqref{eq:ss_chain} and the Planckian bound of Theorem~\ref{thm:planckian_bound} as lower bounds on the thermalization time $\tau_{\delta,\varepsilon}$.
    $^a$Except the static form, which requires only $\|H_S+V\|$.
    $^b$Irrespective of the dynamics: as $1/g$ for the static form and, for $\tau_{\rm SS}$, whenever the fluctuations of $V_0$ do not vanish.
    $^c$Unless the targets become indistinguishable at accuracy $\varepsilon$, as in the RLM (Sec.~\ref{subsec:rlm}).}
    \centering
    \begin{tabular}{lcc}
        \toprule
        & Speed limits & Planckian bound\\
        \midrule
        Cost & Preparation & Information\\
        Rate set by & $H_S+V$ & $\|H_S^{(1)}-H_S^{(2)}\|$\\
        Depends on $\rho_S(0)$ & Yes & No\\
        Requires the dynamics & Yes$^a$ & No\\
        {Depends on the interaction} & {Directly} & {Via the targets}\\
        Vanishes at strong coupling & Yes$^b$ & No$^c$\\
        {\begin{tabular}[t]{@{}l@{}}Vanishes for a single\\Hamiltonian, $\delta=0$\end{tabular}} & No & Yes\\
        \bottomrule
    \end{tabular}
\end{table}

\section{Equilibrium sensitivity of mean-force Gibbs states}
\label{sec:mfg_sensitivity}

This section turns Eq.~\eqref{eq:local_chi} into a computable bound at strong coupling. We express the QFI of the MFG family through equilibrium correlations of the global state and bound it from above and below in terms of a static susceptibility and an equilibrium variance. These two quantities also show when the bound is Planckian. Finally, we optimize over the perturbation $\kappa$.

\subsection{Response of the MFG state without the mean-force Hamiltonian}
\label{subsec:mfg_response}

Evaluating Eq.~\eqref{eq:local_chi} requires the QFI of the MFG family. The MFG state can be written as a Gibbs state of a mean-force Hamiltonian~\cite{Miller2018,trushechkin2022open}, on which previous strong-coupling estimates of the QFI rely, exactly or approximately~\cite{MillerAnders2018EnergyTemperature,Pedram2025FiniteSizeMetrology}, but this Hamiltonian depends on $H_S$ in a way that is generally unknown. We avoid it by differentiating the global Gibbs state instead. Let $\omega_{\rm tot}\equiv\omega(\beta,\bar H_S+H_B+V)$, $\tilde\omega\equiv\tilde\omega(\beta,\bar H_S)$, and $\Delta X\coloneq X-\Tr[\tilde\omega X]\,\mathbb{I}$ for any system operator $X$, which acts on the compound as $X\otimes\mathbb{I}_B$. For each system operator $X$ we define the traceless response operator
\begin{equation}
    R_{X}\coloneq\Tr_B\int_0^1 ds\;\omega_{\rm tot}^{s}\,\Delta {X}\,\omega_{\rm tot}^{1-s}.
    \label{eq:response_operator}
\end{equation}
Since $\delta\kappa$ enters $H_{\rm tot}$ additively, the derivative of the exponential~\cite{metrology_thermalequilibrium,Scandi2025QFIDynamicalNature} gives $\partial_\delta\tilde\omega|_{\delta=0}=-\beta R_\kappa$. Hence, for any system observable $A$, $\partial_\delta\Tr[\tilde\omega A]=-\xi_{A\kappa}$ with
\begin{align}
    \xi_{A\kappa}&\coloneq\beta\Tr[A\,R_\kappa]=\int_0^\beta d\lambda\;{G_{A\kappa}(\lambda)},
    \nonumber\\
    {G_{A\kappa}(\lambda)}&\mathrel{\coloneq}{\Tr}\!\big[\omega_{\rm tot}\,\Delta A(-i\hbar\lambda)\,\Delta\kappa\big],
    \label{eq:static_susceptibility}
\end{align}
where $X(t)=e^{iH_{\rm tot}t/\hbar}Xe^{-iH_{\rm tot}t/\hbar}$, so that $G_{A\kappa}$ is the imaginary-time correlation function. This is the isothermal (static) susceptibility of $A$ with respect to $\kappa$ in the global equilibrium state~\cite{KuboTodaHashitsume1991}, and it is symmetric, $\xi_{A\kappa}=\xi_{\kappa A}$. Equivalently, in terms of the symmetrized equilibrium correlation function $C_{A\kappa}(t)\coloneq\frac12\Tr[\omega_{\rm tot}\{\Delta A(t),\Delta\kappa\}]$ and its spectrum $S_{A\kappa}(\nu)\coloneq\frac{1}{2\pi}\int dt\,e^{i\nu t}C_{A\kappa}(t)$,
\begin{equation}
\begin{split}
    \xi_{A\kappa}&=\beta\int dt\;w_\beta(t)\,C_{A\kappa}(t)\\
    &=\beta\int d\nu\;S_{A\kappa}(\nu)\,\frac{\tanh(\beta\hbar\nu/2)}{\beta\hbar\nu/2},
\end{split}
    \label{eq:susceptibility_correlations}
\end{equation}
with $w_\beta(t)=\frac{2}{\pi\beta\hbar}\ln\coth\frac{\pi|t|}{2\beta\hbar}$ a normalized kernel of width $\sim\hbar\beta$, the \emph{Planckian window}, whose Fourier transform is the \emph{Planckian filter} $\tanh(\beta\hbar\nu/2)/(\beta\hbar\nu/2)$~\cite{Kubo1966FDT,MengShi2026ThermodynamicUncertainty}. The sensitivity of the MFG state is thus fixed by equilibrium correlations of system observables, averaged over a window of the order of the Planckian time~\cite{Pappalardi2022QuantumBounds}. It requires neither the mean-force Hamiltonian nor the relaxation dynamics. Note that the system observables in $C_{A\kappa}$ evolve with $H_{\rm tot}$, so that $S_{\kappa\kappa}$ is not the spectral density of the bath. In electron transport in bad metals, the same Planckian filter appears as a quantum detailed-balance factor and yields an apparent Planckian scattering rate when current fluctuations decay on times much shorter than $\hbar/k_BT$~\cite{Fratini2026StrangeMetal}. Derivations are given in Apps.~\ref{app:mfg_response} and \ref{app:spectral}.

\subsection{Quantum Fisher information of the MFG state}
\label{subsec:mfg_qfi}

\begin{proposition}[QFI of the MFG state]\label{prop:mfg_qfi}
The QFI entering Eq.~\eqref{eq:local_chi} is
\begin{equation}
    \mathcal F^{\rm th}_\kappa(\beta,\bar H_S)=\max_A\frac{\xi_{A\kappa}^2}{\operatorname{Var}_{\tilde\omega}(A)},
    \label{eq:qfi_variational}
\end{equation}
where the maximum runs over system observables, $\operatorname{Var}_{\tilde\omega}(A)=\Tr\big[\tilde\omega A^2\big]-\Tr\big[\tilde\omega A\big]^2$, and is attained by the symmetric logarithmic derivative (SLD) of the MFG family. Moreover,
\begin{equation}
    \frac{\xi_{\kappa\kappa}^2}{\operatorname{Var}_{\tilde\omega}(\kappa)}
    \leq\mathcal F^{\rm th}_\kappa
    \leq\beta\,\xi_{\kappa\kappa}
    \leq\beta^2\operatorname{Var}_{\tilde\omega}(\kappa)
    \leq\frac{\beta^2\|\kappa\|^2}{4}.
    \label{eq:qfi_hierarchy}
\end{equation}
\end{proposition}

Equation~\eqref{eq:qfi_variational} is the variational form of the QFI~\cite{Paris2009QuantumEstimation,BraunsteinCaves1994} combined with Eq.~\eqref{eq:static_susceptibility}. The first inequality in Eq.~\eqref{eq:qfi_hierarchy} follows by choosing $A=\kappa$. For the global Gibbs state, the second and third are, respectively, the bound of the QFI by the Kubo--Mori variance and of the latter by the variance~\cite{MillerAnders2018EnergyTemperature}; the full chain is stated for Gibbs states in Ref.~\cite{MengShi2026ThermodynamicUncertainty}. They carry over to the MFG family because the QFI is monotone under the partial trace, $\mathcal F^{\rm th}_\kappa\leq\mathcal F[\omega(\beta,H_{\rm tot})]$, while $\xi_{\kappa\kappa}$ and $\operatorname{Var}_{\tilde\omega}(\kappa)$ take the same values in the global state. The last inequality is the variance bound of Refs.~\cite{Boixo2007GeneralizedLimits,metrology_thermalequilibrium}.

For $\|\kappa\|=1$, any choice of $\kappa$ and $A$ thus gives a valid bound, $\tau\geq\hbar|\xi_{A\kappa}|/\sqrt{\operatorname{Var}_{\tilde\omega}(A)}$. The simplest one, $A=\kappa$, reads
\begin{equation}
    \tau\;\geq\;\frac{\hbar\,\xi_{\kappa\kappa}}{\sqrt{\operatorname{Var}_{\tilde\omega}(\kappa)}}
    =\tau_{\rm Pl}\,r_\kappa\,\sqrt{\operatorname{Var}_{\tilde\omega}(\kappa)},
    \label{eq:self_response_bound}
\end{equation}
and requires only one static susceptibility and one equilibrium variance. The factor
\begin{align}
    r_\kappa&\coloneq\frac{\xi_{\kappa\kappa}}{\beta\operatorname{Var}_{\tilde\omega}(\kappa)}
    =\frac{\int dt\,w_\beta(t)\,C_{\kappa\kappa}(t)}{C_{\kappa\kappa}(0)}
    \nonumber\\
    &{=\frac1\beta\int_0^\beta d\lambda\,\frac{G_{\kappa\kappa}(\lambda)}{G_{\kappa\kappa}(0)}}\in(0,1]
    \label{eq:r_kappa}
\end{align}
is the average of the normalized autocorrelation of $\kappa$ over the Planckian window $w_\beta$ or, equivalently, over imaginary time, with $G_{\kappa\kappa}(0)=G_{\kappa\kappa}(\beta)=\operatorname{Var}_{\tilde\omega}(\kappa)$. Since $G_{\kappa\kappa}(\lambda)$ is convex and symmetric about $\lambda=\beta/2$, it is minimal there, and
\begin{align}
    {r_\kappa}&\mathrel{\geq}{\frac{G_{\kappa\kappa}(\beta/2)}{\operatorname{Var}_{\tilde\omega}(\kappa)},}
    \nonumber\\
    {G_{\kappa\kappa}(\beta/2)}&\mathrel{=}{\int d\nu\,S_{\kappa\kappa}(\nu)\,{\rm sech}(\beta\hbar\nu/2),}
    \label{eq:r_floor}
\end{align}
where $G_{\kappa\kappa}(\beta/2)$ retains only the fluctuations slower than about $k_BT/\hbar$. We use this bound in Sec.~\ref{sec:heisenberg}. In the global Gibbs state, $r_\kappa$ is the ratio of the Kubo--Mori to the variance metric~\cite{Tonchev2021MonotoneMetrics}. Taking square roots in Eq.~\eqref{eq:qfi_hierarchy} and dividing by $\beta$ brackets the optimal bound for a given $\kappa$, in units of $\tau_{\rm Pl}$, between quantities that only involve $\xi_{\kappa\kappa}$ and $\operatorname{Var}_{\tilde\omega}(\kappa)$:
\begin{equation}
    r_\kappa\sqrt{\operatorname{Var}_{\tilde\omega}(\kappa)}
    \leq\frac{\sqrt{\mathcal F^{\rm th}_\kappa}}{\beta}
    \leq\sqrt{r_\kappa}\,\sqrt{\operatorname{Var}_{\tilde\omega}(\kappa)}
    \leq\frac12.
    \label{eq:r_bracket}
\end{equation}
For a given $\kappa$, Eq.~\eqref{eq:r_bracket} is a diagnostic: no choice of the observable $A$ can improve Eq.~\eqref{eq:self_response_bound} by more than a factor $r_\kappa^{-1/2}$. Maximizing its upper end over $\kappa$ gives instead a ceiling for the whole problem, $\widetilde\chi\leq\max_{\,{\|\kappa\|=1}}\sqrt{\xi_{\kappa\kappa}/\beta}$: when it is small, no perturbation and no observable yields a strong bound.
The limiting value $r_\kappa=1$ requires $[\kappa,H_{\rm tot}]=0$ (App.~\ref{app:spectral}). Such a $\kappa$ is then a constant of motion, so that $\Tr[\rho_S(\tau)\kappa]$ keeps its initial value for every member of the family, whereas the target value $\Tr[\tilde\omega\kappa]$ does depend on $\delta$. Definition~\ref{def:thermalization} therefore cannot be met in the locally exact limit, and this limiting case is excluded. The ceiling $1/2$ in Eq.~\eqref{eq:r_bracket} is hence never reached. It is approached only when $\operatorname{Var}_{\tilde\omega}(\kappa)\to1/4$ and the fluctuations of $\kappa$ remain correlated over times much longer than $\tau_{\rm Pl}$, as for a weakly coupled level at resonance (Sec.~\ref{subsec:rlm}).

Equation~\eqref{eq:r_bracket} also answers whether the bound remains Planckian. Both of its ends depend on two equilibrium properties of the perturbation: the size of its fluctuations, $\sqrt{\operatorname{Var}_{\tilde\omega}(\kappa)}\leq\frac12$, and how long these fluctuations remain correlated, measured by $r_\kappa$. Since $w_\beta$ has a width of order $\hbar\beta$, $r_\kappa\simeq1$ when $C_{\kappa\kappa}(t)$ stays close to $C_{\kappa\kappa}(0)$ over times of order $\tau_{\rm Pl}$, and $r_\kappa\ll1$ when it decays much faster. This leads to the following criterion.

\begin{criterion}
The thermalization bound is Planckian when some perturbation $\kappa$ of $H_S$ fluctuates appreciably in equilibrium, with $\sqrt{\operatorname{Var}_{\tilde\omega}(\kappa)}$ comparable with $1/2$, and its fluctuations remain correlated over the Planckian time $\hbar/k_BT$. It falls below the Planckian scale when, for every perturbation, the fluctuations are small or decorrelate on times much shorter than $\hbar/k_BT$.
\end{criterion}

The criterion is clearest in frequency. By Eq.~\eqref{eq:susceptibility_correlations}, $r_\kappa$ is the average of the Planckian filter $f(\beta\hbar\nu)$ over the normalized spectrum $S_{\kappa\kappa}(\nu)/C_{\kappa\kappa}(0)$. The filter is close to $1$ for $|\nu|\ll k_BT/\hbar$ and reduces to $2k_BT/(\hbar|\nu|)$ for $|\nu|\gg k_BT/\hbar$, so $r_\kappa$ measures the effective fraction of the fluctuations of $\kappa$ at frequencies $|\nu|\lesssim k_BT/\hbar$. When this fraction is of order one, Eq.~\eqref{eq:self_response_bound} gives the Planckian scale, $\tau\gtrsim\tau_{\rm Pl}\sqrt{\operatorname{Var}_{\tilde\omega}(\kappa)}$. Its upper limit, $r_\kappa=1$, would require all the spectral weight at $\nu=0$, i.e., a conserved perturbation, which was excluded below Eq.~\eqref{eq:r_bracket} because its expectation value stays at its initial value while the target varies with $H_S$. The Planckian constraint is therefore carried by fluctuations that are slow but not frozen. When instead the spectrum lies at $|\nu|\gg k_BT/\hbar$, $r_\kappa\simeq2k_BT/(\hbar\nu_0)\ll1$, with $\nu_0$ the harmonic mean of $|\nu|$ over the normalized spectrum, and Eq.~\eqref{eq:self_response_bound} gives $\tau\gtrsim2\sqrt{\operatorname{Var}_{\tilde\omega}(\kappa)}/\nu_0$, set by the fluctuation time scale $1/\nu_0$ rather than by $\tau_{\rm Pl}$, as for gapped ground states~\cite{metrology_thermalequilibrium}. The criterion is qualitative, since correlations that decay on several time scales, some shorter and some longer than $\hbar\beta$, yield intermediate values of $r_\kappa$. Sections~\ref{sec:fermionic}--\ref{sec:heisenberg} illustrate both regimes.

\subsection{Optimal perturbations and explicit bounds}
\label{subsec:optimal_perturbations}

The criterion requires some perturbation with large and slowly decorrelating equilibrium fluctuations. It remains to identify the perturbation that yields the strongest bound, which Ref.~\cite{Abiuso2025PlanckianBound} addressed through trial families. The optimization over perturbations in Eq.~\eqref{eq:local_chi} can be restricted, without loss of generality, to projectors. The argument uses three elementary properties of $\sqrt{\mathcal F^{\rm th}_\kappa}$. (a)~\emph{Homogeneity:} since $\partial_\delta\tilde\omega=-\beta R_\kappa$ is linear in $\kappa$, $\sqrt{\mathcal F^{\rm th}_{c\kappa}}=|c|\sqrt{\mathcal F^{\rm th}_\kappa}$. The maximum over $\|\kappa\|=1$ in Eq.~\eqref{eq:local_chi} therefore equals the maximum over $\|\kappa\|\leq1$. (b)~\emph{Shift invariance:} $\kappa\to\kappa+c\,\mathbb{I}$ adds a constant to $H_{\rm tot}$, which leaves $\omega_{\rm tot}$, $\tilde\omega$, and $\mathcal F^{\rm th}_\kappa$ unchanged, and does not change $\|\kappa\|$. Shifting $\kappa$ so that $\lambda_{\min}(\kappa)=0$, the set $\|\kappa\|\leq1$ becomes the set of operators with eigenvalues in $[0,1]$, i.e., $\{0\leq\kappa\leq\mathbb{I}\}$. (c)~\emph{Convexity:} by Eq.~\eqref{eq:qfi_variational}, $\sqrt{\mathcal F^{\rm th}_\kappa}=\max_A|\xi_{A\kappa}|/\sqrt{\operatorname{Var}_{\tilde\omega}(A)}$ is a maximum of absolute values of linear functions of $\kappa$, hence convex. Writing $\kappa=\sum_i\lambda_i|i\rangle\langle i|$, for a fixed eigenbasis $\sqrt{\mathcal F^{\rm th}_\kappa}$ is a convex function of $(\lambda_1,\dots,\lambda_d)\in[0,1]^d$. Its maximum is thus attained at a vertex of this cube, $\lambda_i\in\{0,1\}$, i.e., at a projector $P$. Excluding the trivial cases $P=0,\mathbb{I}$, for which $\mathcal F^{\rm th}_P=0$, all projectors have $\|P\|=1$, and the bound of Eq.~\eqref{eq:local_chi} becomes
\begin{equation}
    \tau\;\geq\;\tau_{\rm Pl}\,\widetilde\chi(\bar H_S)
    =\hbar\max_P\sqrt{\mathcal F^{\rm th}_P}
    =\frac{\tau_{\rm Pl}}{2}\max_A\frac{\|R_A\|_1}{\sqrt{\operatorname{Var}_{\tilde\omega}(A)}},
    \label{eq:projector_bound}
\end{equation}
with $\|R\|_1=\Tr|R|$. The second form is a constructive version of the same statement. For a fixed observable $A$, $\xi_{A\kappa}=\beta\Tr[\kappa R_A]$ is linear in $\kappa$, and its maximum over $0\leq\kappa\leq\mathbb{I}$ is the sum of the positive eigenvalues of $R_A$, attained at the projector $P_+(R_A)$ onto the corresponding eigenspace. Since $R_A$ is traceless, this sum equals $\|R_A\|_1/2$. The optimal perturbation and the optimal measurement are thus related self-consistently: $\kappa^*=P_+(R_{A^*})$, with $A^*$ the SLD for $\kappa^*$. Physically, $\kappa=P$ raises by $\delta$ the energy of the states in the range of $P$ relative to its complement; when $[P,\bar H_S]=0$, this is a rigid shift of a group of levels of $\bar H_S$. The most demanding local tests of thermalization are of this type. Since $\sqrt{\xi_{\kappa\kappa}}$ is also convex, homogeneous and shift-invariant in $\kappa$, the same argument restricts the ceiling below Eq.~\eqref{eq:r_bracket} to projectors, $\widetilde\chi\leq\max_P\sqrt{\xi_{PP}/\beta}$. In practice, expanding in a basis of traceless system observables $\{A_i\}$, $\kappa=\sum_ik_iA_i+c\,\mathbb{I}$, Eq.~\eqref{eq:qfi_variational} becomes $\mathcal F^{\rm th}_\kappa=\beta^2k^\top M\Sigma^{-1}Mk$, with $M_{ij}=\xi_{A_iA_j}/\beta$ and $\Sigma_{ij}=\frac12\Tr[\tilde\omega\{\Delta A_i,\Delta A_j\}]$.

The projector reduction and the basis expansion yield explicit bounds in several cases. (i)~At weak coupling, $g\to0$, $\omega_{\rm tot}\to\omega(\beta,\bar H_S)\otimes\omega(\beta,H_B)$ and Eq.~\eqref{eq:qfi_variational} reduces to the QFI of the bare Gibbs state~\cite{metrology_thermalequilibrium}, recovering Ref.~\cite{Abiuso2025PlanckianBound}. For $[\kappa,\bar H_S]=0$ this QFI is $\beta^2\operatorname{Var}(\kappa)$~\cite{ZanardiCamposVenutiGiorda2007BuresThermal,metrology_thermalequilibrium}, and the projector reduction shows that the diagonal lower bound of Ref.~\cite{Abiuso2025PlanckianBound}, the maximum of $\sqrt{p_I(1-p_I)}$ over sets $I$ of eigenstates of $\bar H_S$ with total population $p_I$, is the optimum among such perturbations. (ii)~If symmetries keep the MFG family diagonal in a fixed basis $\{|i\rangle\}$, then $\mathcal F^{\rm th}_\kappa=\sum_i\xi_{\Pi_i\kappa}^2/p_i$, with $\Pi_i=|i\rangle\langle i|$ and $p_i=\Tr[\tilde\omega\,\Pi_i]$. For two levels and $\kappa=P$, the projector onto one of them, the first inequality in Eq.~\eqref{eq:qfi_hierarchy} is tight and
\begin{equation}
    \tau\;\geq\;\frac{\hbar\,|\partial_\delta p|}{\sqrt{p(1-p)}},
    \qquad p=\Tr[\tilde\omega P],
    \label{eq:two_level_bound}
\end{equation}
which only requires the equilibrium population of $P$ and its response to a shift of the corresponding energy. This applies, e.g., to the occupation of a single fermionic level, $P=d^\dagger d$, as in the resonant-level model discussed below. (iii)~For a qubit, write $\tilde\omega=\frac12(\mathbb{I}+\bm b\cdot\bm\sigma)$ and $\kappa=\frac12\bm n\cdot\bm\sigma$ with $|\bm n|=1$, so that $\|\kappa\|=1$ and all projectors are covered. The Bloch vector responds as $\partial_\delta\bm b=J\bm n$, with the symmetric static response matrix $J_{ij}=-\xi_{\sigma_i\sigma_j}/2$. With $A_i=\sigma_i$ one has $\Sigma=\mathbb{I}-\bm b\bm b^\top$, and $\mathcal F^{\rm th}_\kappa=|\partial_\delta\bm b|^2+(\bm b\cdot\partial_\delta\bm b)^2/(1-|\bm b|^2)$~\cite{Zhong2013BlochFisher}, so that
\begin{equation}
\widetilde\chi=\frac1\beta\sqrt{\lambda_{\max}\!\Big[J\Big(\mathbb{I}+\frac{\bm b\,\bm b^\top}{1-|\bm b|^2}\Big)J\Big]}.
    \label{eq:qubit_chi}
\end{equation}
Reference~\cite{Abiuso2025PlanckianBound} optimized a one-parameter family of qubit perturbations for Gibbs targets; Eq.~\eqref{eq:qubit_chi} holds for any MFG Bloch vector and static response matrix. If $\bm b$ is an eigenvector of $J$, e.g., when $H_{\rm tot}$ is invariant under a $\pi$ rotation of the qubit about the axis of $\bm b$ (combined with a unitary on $B$), the optimization separates into a longitudinal (population) and a transverse (coherence) channel,
\begin{equation}
    \widetilde\chi=\frac1\beta\max\Big\{\frac{|J_\parallel|}{\sqrt{1-|\bm b|^2}},\,|J_{\perp,1}|,\,|J_{\perp,2}|\Big\},
    \label{eq:qubit_channels}
\end{equation}
where $J_\parallel$ and $J_{\perp,k}$ are the eigenvalues of $J$ along and orthogonal to $\bm b$. A $U(1)$ symmetry about the same axis further imposes $J_{\perp,1}=J_{\perp,2}$. In each channel the SLD is proportional to $\kappa$ up to a multiple of $\mathbb{I}$, so Eq.~\eqref{eq:self_response_bound} is exact (App.~\ref{app:qubit}). At weak coupling, with $\bar H_S=\frac{\Delta}{2}\sigma_z$, the two channels give $\frac12{\rm sech}(\beta\Delta/2)$ and $\tanh(\beta\Delta/2)/(\beta\Delta)$; the transverse one is always larger, in agreement with Ref.~\cite{metrology_thermalequilibrium}.

\section{Strongly coupled fermionic models: loss and protection of equilibrium information}
\label{sec:fermionic}

We first test the bound in noninteracting fermionic impurity models, i.e., quantum dots tunnel-coupled to a metallic lead.
Their MFG states and reduced dynamics are known exactly at arbitrary coupling, which makes them an ideal benchmark: the bound, which requires only equilibrium information, can be compared with exact thermalization times and with open-system speed limits.
For each model, we identify the optimal perturbation and the observable that recovers it, and then compare the finite-error bound with the exact time $\tau_\varepsilon$ at which a common dynamics brings a family of Hamiltonians within accuracy $\varepsilon$ of their targets.

The two models realize the two regimes of the criterion. In the resonant-level model (RLM), strong coupling makes the level occupation decorrelate on times much shorter than $\tau_{\rm Pl}$: the bound loses its Planckian character and becomes comparable to the relaxation time $\hbar/g^2$. In the double quantum dot (DQD), a dark mode decoupled from the reservoir keeps slow equilibrium fluctuations, so that the bound remains Planckian at every coupling, while thermalization slows down. In both models, speed limits based on the interaction energy vanish as the bandwidth of the reservoir grows, whereas the bound and $\tau_\varepsilon$ do not.

\subsection{Model and exact solution}
\label{subsec:fermionic_model}

We consider $N$ fermionic levels $d_n$ coupled to a common reservoir,
\begin{align}
    H_S&=\sum_{n=1}^{N}\epsilon_n\,d_n^\dagger d_n,
    \qquad
    H_B=\sum_k\Omega_k\,b_k^\dagger b_k,
    \nonumber\\
    V&=g\sum_{n=1}^{N}\sum_k\left(t_k\,d_n^\dagger b_k+t_k^*\,b_k^\dagger d_n\right),
    \label{eq:fermionic_hamiltonian}
\end{align}
with all energies measured from the chemical potential of the reservoir. All levels couple to the same reservoir channel with equal amplitudes, as for quantum dots tunnel-coupled to a common metallic lead.
The reservoir enters through its spectral density, which we take to be Lorentzian~\cite{Rolandi2023finitetimelandauer},
\begin{equation}
    \mathcal J(\Omega)=2\pi\sum_k|t_k|^2\,\delta(\Omega-\Omega_k)=\frac{\Lambda^2}{\Lambda^2+\Omega^2}.
    \label{eq:lorentzian}
\end{equation}
Thus $g^2\mathcal J(\Omega)$ is the energy-dependent level broadening, and $g^2$, the hybridization energy at the Fermi level, measures the coupling strength throughout.

We work in the wide-band regime, in which $\Lambda$ is finite but much larger than $g^2$, $|\epsilon_n|$, and $k_BT$. This regime describes all dynamics on time scales long compared with $\hbar/\Lambda$, which include the thermalization times considered here, but not the initial transient at $t\lesssim\hbar/\Lambda$~\cite{Rolandi2023finitetimelandauer}. Within it, the model is exactly solvable at arbitrary coupling: both the MFG state and the reduced dynamics are known in closed form, without weak-coupling or Markovian approximations.

The total Hamiltonian is quadratic and conserves the particle number. For an initially factorized state $\rho_S(0)\otimes\omega(\beta,H_B)$ with a number-conserving Gaussian $\rho_S(0)$, the reduced state therefore remains Gaussian and is fully determined by its correlation matrix $C_{mn}(t)=\Tr[\rho_S(t)\,d_n^\dagger d_m]$~\cite{suraceFermionicGaussianStates2022}. Its relaxation is governed by the matrix
\begin{equation}
    \Gamma=-\frac{i}{\hbar}\,h-\frac{g^2}{2\hbar}\,\bm u\bm u^{\rm T},
    \qquad
    \bm u=(1,\dots,1)^{\rm T},
    \label{eq:damping_matrix}
\end{equation}
with $h={\rm diag}(\epsilon_1,\dots,\epsilon_N)$, in the sense that the memory of the initial state decays as $D(t)C(0)D(t)^{\dagger}$ with $D(t)=e^{\Gamma t}$ (Appendix~\ref{app:fermionic}). The collective term $\bm u \bm u^{\rm T}$ is the origin of the bright and dark modes discussed below. Whenever all eigenvalues of $\Gamma$ have negative real parts, $C(t)$ converges to the correlation matrix of the MFG state (Appendix~\ref{app:fermionic}; see Ref.~\cite{Rolandi2023finitetimelandauer} for a single level),
\begin{equation}
    C^{\rm th}=\frac12\,\mathbb{I}-\frac{1}{2\pi i}\left[\psi\!\left(\frac12\,\mathbb{I}-\frac{\beta\hbar}{2\pi}\Gamma\right)-{\rm h.c.}\right],
    \label{eq:mfg_correlation}
\end{equation}
where $\psi$ is the digamma function. In these models the stationary state thus coincides with the strong-coupling thermal target $\tilde\omega(\beta,H_S)$, and both the targets and their QFI follow from $C^{\rm th}$ and its derivative~\cite{carolloSymmetricLogarithmicDerivative2018}.

For the finite-error tests we fix a thermalization task and vary only the coupling. The task consists of a family of system Hamiltonians along a fixed direction $\hat\kappa$,
\begin{equation}
    H_S(x)=\bar H_S+x\,\hat\kappa,\qquad \|\hat\kappa\|=1,\qquad |x|\leq\delta,
    \label{eq:hamiltonian_family}
\end{equation}
a common factorized initial state, and an accuracy $\varepsilon$. In all models we take $\delta=2\pi k_BT$. For a single weakly coupled level at resonance, the thermal occupation then goes from nearly $1$ at $x=-\delta$ to nearly $0$ at $x=\delta$, so that the extreme targets are nearly orthogonal. In the other regimes and models, the same fixed $\delta$ tests whether the equilibrium state still resolves Hamiltonians on this scale. Its thermalization time, which we denote $\tau_\varepsilon$ since $\delta$ is fixed, is the thermalization time of Eq.~\eqref{eq:thermalization_time} with the ball replaced by this family,
\begin{equation}
    \tau_\varepsilon\coloneq\min\Big\{t\geq0:\ \max_{|x|\leq\delta}D\big(\rho_S(t,x),\tilde\omega_x\big)\leq\varepsilon\Big\},
    \label{eq:tau_eps}
\end{equation}
with $\rho_S(t,x)\equiv\rho_S(t,H_S(x))$ and $\tilde\omega_x\equiv\tilde\omega(\beta,H_S(x))$. Since the proof of Theorem~\ref{thm:planckian_bound} compares Hamiltonians pairwise, the theorem applies to this family and gives $\tau_\varepsilon\geq\tau_{\rm Pl}\chi_\delta$, where $\chi_\delta$ is the factor $\chi$ of Eq.~\eqref{eq:chi_def} with the maximum restricted to pairs of members, for which $\|H_S(x_1)-H_S(x_2)\|=|x_1-x_2|$. Since Bures angles never exceed $\pi/2$, $\chi_\delta$ can be nonzero only for $\varepsilon<\varepsilon_{\max}=\pi/4$, and we quote accuracies as fractions of $\varepsilon_{\max}$, using the same values in all models. For the whole ball $\mathcal B_\delta(\bar H_S)$, both the thermalization time $\tau_{\delta,\varepsilon}$ and the bound can only be larger, since the family is contained in it. In the locally exact limit, the bound for the ball is set by the most sensitive direction, which we identify in Secs.~\ref{subsec:rlm} and \ref{subsec:dqd}, although it need not be the most useful one for comparison with $\tau_\varepsilon$ (Sec.~\ref{subsec:dqd}).

\begin{figure*}[t]
    \centering
    \includegraphics[width=\textwidth]{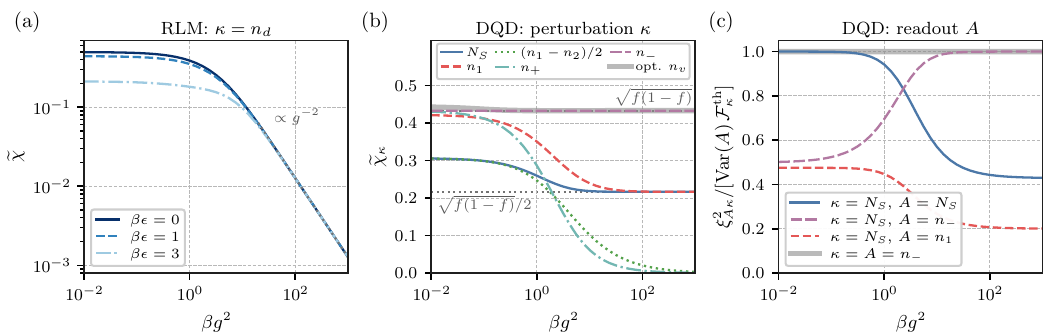}
    \caption{\textbf{Locally exact constraints and their recovery by measurements.}
    (a)~RLM: $\widetilde\chi=\sqrt{\mathcal F^{\rm th}_\kappa}/\beta$ for the level shift $\kappa=n_d$, the only admissible perturbation, at three level energies. The occupation readout recovers this value exactly, as Eq.~\eqref{eq:rlm_chi} shows. Dotted line: strong-coupling limit $4/(\pi\beta g^2)$.
    (b)~DQD at $\beta\bar\epsilon=1.1$ and $\beta\Delta=0.2$: $\widetilde\chi_\kappa=\sqrt{\mathcal F^{\rm th}_\kappa}/\beta$ for several quadratic perturbations, and its maximum over all of them, attained at a single-mode occupation $n_v$ (thick gray). The interdot tunneling $(d_1^\dagger d_2+{\rm h.c.})/2$ coincides with the common shift within $0.3\%$ and is omitted. Dotted lines: strong-coupling limits $\sqrt{f(1-f)}$ and $\sqrt{f(1-f)}/2$, with $f=f_\beta(\bar\epsilon)$.
    (c)~Fraction of $\mathcal F^{\rm th}_\kappa$ recovered by the mean response of an observable $A$, as in Eq.~\eqref{eq:qfi_variational}, for the common shift and for the dark-mode perturbation. Wide-band MFG states.}
    \label{fig:local_sensitivity}
\end{figure*}

\subsection{Resonant-level model: strong coupling erases equilibrium information}
\label{subsec:rlm}

For a single level, $H_S=\epsilon\,n_d$ with $n_d=d^\dagger d$. Since physical fermionic Hamiltonians and observables conserve the fermion parity, and the parity-even operators of a single mode are spanned by $\mathbb{I}$ and $n_d$, the only admissible perturbation is, up to a constant, the level shift $\kappa=n_d$, a projector with $\|n_d\|=1$. Number conservation keeps the MFG state diagonal in $\{|0\rangle,|1\rangle\}$, so case (ii) of Sec.~\ref{subsec:optimal_perturbations} applies: measuring the occupation recovers the full QFI, and Eq.~\eqref{eq:two_level_bound} holds with
\begin{align}
    \widetilde\chi&=\frac{|\partial_\epsilon p_{\rm th}|}{\beta\sqrt{p_{\rm th}(1-p_{\rm th})}},
    \nonumber\\
    p_{\rm th}&=\frac12-\frac1\pi\,{\rm Im}\,\psi\!\left(\frac12+\frac{\beta}{2\pi}\Big(\frac{g^2}{2}+i\epsilon\Big)\right).
    \label{eq:rlm_chi}
\end{align}
The locally exact bound of the RLM is thus fixed exactly by the equilibrium occupation and its static response.

Figure~\ref{fig:local_sensitivity}(a) shows how this sensitivity depends on the coupling. At weak coupling, $p_{\rm th}\to f_\beta(\epsilon)$, with $f_\beta(\epsilon)=(1+e^{\beta\epsilon})^{-1}$ the Fermi function, and $\widetilde\chi\to\frac12{\rm sech}(\beta\epsilon/2)$. At $\epsilon=0$ this equals $1/2$, the largest value allowed by Eq.~\eqref{eq:r_bracket}, so the weakly coupled resonant level approaches the largest locally exact constraint that any perturbation can impose, $\tau\geq\tau_{\rm Pl}/2$. As $g^2$ increases, the level is broadened over an energy $g^2$, its occupation approaches $1/2$ independently of $\epsilon$, and $\widetilde\chi\to4/(\pi\beta g^2)$. The locally exact constraint then becomes $\tau\geq4\hbar/(\pi g^2)$, independent of temperature: the occupation fluctuates at frequencies of order $g^2/\hbar\gg k_BT/\hbar$, in agreement with the criterion of Sec.~\ref{subsec:mfg_qfi}. The surviving constraint is set by the coupling itself and is comparable to the occupation relaxation time $\hbar/g^2$. Strong coupling thus removes the Planckian character of the bound. Since the level shift and the occupation are the only nontrivial perturbation and observable of a single level, this loss reflects the equilibrium state itself, not a poor choice of perturbation or observable.

\textit{Finite-error test}. We take the family $H_S(x)=x\,n_d$ with $|x|\leq\delta=2\pi k_BT$, centered at resonance, and the common preparation $p(0)=1/2$.
Since the targets are diagonal in $\{|0\rangle,|1\rangle\}$, the Bures angle between two of them is $D(\tilde\omega_x,\tilde\omega_y)=|\theta_x-\theta_y|$, with $\theta_x=\arcsin\sqrt{p_{\rm th}(x)}$ monotonic in $x$. Particle--hole symmetry gives $\theta_{-x}=\pi/2-\theta_x$, so the preparation, with $\theta=\pi/4$, is the geodesic midpoint of the extreme targets, and $\max_{|x|\leq\delta}D(\rho_S(0),\tilde\omega_x)=\frac12D(\tilde\omega_{-\delta},\tilde\omega_\delta)$. All members thus start within $\varepsilon$ of their targets if and only if $D(\tilde\omega_{-\delta},\tilde\omega_\delta)\leq2\varepsilon$, which is also the condition for $\chi_\delta=0$. The thermalization time and the bound therefore vanish at the same coupling.

\begin{figure*}[t]
    \centering
    \includegraphics[width=\textwidth]{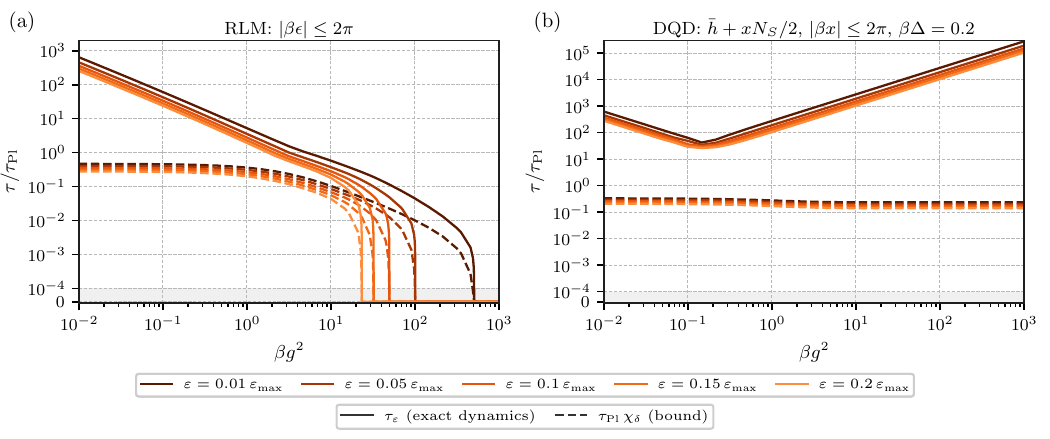}
    \caption{\textbf{Thermalization times and bounds for fixed Hamiltonian families.}
    Exact time $\tau_\varepsilon$ at which all members of the family are within $\varepsilon$ of their targets, defined in Eq.~\eqref{eq:tau_eps} (solid), and bound $\tau_{\rm Pl}\chi_\delta$ of Theorem~\ref{thm:planckian_bound} (dashed) versus coupling, for five accuracies.
    (a)~RLM: $H_S(\epsilon)=\epsilon\,n_d$ with $|\beta \epsilon|\leq2\pi$ and initial occupation $p(0)=1/2$.
    (b)~DQD: common shift $h={\rm diag}(1.2,1.0)\,k_BT+x\,\mathbb{I}/2$, i.e., $\kappa=N_S/2$ at fixed $\beta\Delta=0.2$, with $|\beta x|\leq2\pi$ and initial occupations $(n_1,n_2)=(1,0)$. 
    The vertical axis is linear below $10^{-4}$, so that exact zeros remain visible. Wide-band dynamics.}
    \label{fig:finite_error}
\end{figure*}

Figure~\ref{fig:finite_error}(a) shows the result. At weak coupling, $\tau_\varepsilon$ decreases as $\hbar/g^2$, while the bound settles on a plateau set by $\tau_{\rm Pl}$, lower for larger $\varepsilon$. A stronger coupling accelerates relaxation, but it also broadens the level over $g^2$, so that a shift of order $g^2$, rather than $k_BT$, is needed to change its occupation appreciably. For $g^2\gg\delta$, all targets of the family thus cluster around $p_{\rm th}=1/2$. Both $\tau_\varepsilon$ and the bound then vanish beyond a threshold coupling $g_c^2$ that grows linearly with $\delta$ in the strong-coupling regime, so widening the family only postpones the threshold. For $g^2<g_c^2$, the bound stays within a factor of a few of $\tau_\varepsilon$. The locally exact bound, in contrast, never vanishes. It approaches $\tau\geq4\hbar/(\pi g^2)$, below the Planckian scale but scaling with the coupling as the relaxation time $\hbar/g^2$.

\subsection{Double quantum dot: a dark mode protects equilibrium information}
\label{subsec:dqd}

For two levels, write $\bar\epsilon=(\epsilon_1+\epsilon_2)/2$ and $\Delta=\epsilon_1-\epsilon_2$. In terms of the bright and dark modes $d_\pm=(d_1\pm d_2)/\sqrt2$,
\begin{align}
    H_S&=\bar\epsilon\,(n_++n_-)+\frac{\Delta}{2}\left(d_+^\dagger d_-+d_-^\dagger d_+\right),
    \nonumber\\
    V&=\sqrt2\,g\sum_k\left(t_k\,d_+^\dagger b_k+t_k^*\,b_k^\dagger d_+\right),
    \label{eq:bright_dark}
\end{align}
with $n_\pm=d_\pm^\dagger d_\pm$. The reservoir couples only to the bright mode, with an enhanced amplitude, so $[n_-,V]=0$. The dark mode relaxes only through the detuning, which mixes it with the bright mode. Accordingly, the eigenvalues of $\Gamma$ are $-i\bar\epsilon/\hbar-g^2/(2\hbar)\pm\sqrt{g^4-\Delta^2}/(2\hbar)$. For $g^2\gg|\Delta|$ their real parts separate into a fast rate $\Gamma_{\rm fast}=g^2/\hbar$ and a slow rate $\Gamma_{\rm slow}=\Delta^2/(4\hbar g^2)$, so the dark occupation relaxes on the time scale $2\hbar g^2/\Delta^2$, which grows with the coupling. At $\Delta=0$, $n_-$ commutes with $H_{\rm tot}$ and is exactly conserved. The reduced state then retains its initial dark occupation and does not relax to the MFG state~\cite{Ravell,Wei2023MultipleSteadyStates}: this is the conservation obstruction of Sec.~\ref{subsec:mfg_qfi}, with $r_{n_-}=1$. We therefore take $\beta\Delta=0.2$, for which every mode relaxes, albeit slowly at strong coupling.

The equilibrium state reflects the same structure. At strong coupling, the bright mode is broadened over an energy $g^2$ and its occupation is pinned at $1/2$. The dark mode, whose width $\Delta^2/(2g^2)$ vanishes, retains the thermal occupation $f_\beta(\bar\epsilon)$, and the bright--dark coherence vanishes. The MFG state therefore keeps the information on $\bar\epsilon$ that the RLM loses.

\paragraph*{Optimal perturbation.} The controls of a DQD are the gate voltages, which shift $\epsilon_1$ and $\epsilon_2$, and the interdot tunneling. They generate quadratic number-conserving perturbations $\kappa=\sum_{mn}k_{mn}d_m^\dagger d_n$, with a Hermitian $2\times2$ matrix $k$ that is added to $h$ in Eqs.~\eqref{eq:damping_matrix} and \eqref{eq:mfg_correlation}. Their spectral seminorm on Fock space is $\|\kappa\|=\sum_i|\,{\lambda_i}(k)|$; for instance, the normalized common shift is $N_S/2$, since $\|N_S\|=2$ for $N_S=n_1+n_2$. Since the extreme points of the set $\|\kappa\|\leq1$ are, up to sign, the single-mode occupations $n_v=d_v^\dagger d_v$, with $d_v=\sum_nv_n^*d_n$ and $|v|=1$, the convexity argument of Sec.~\ref{subsec:optimal_perturbations} places the optimum at some $n_v$, and only the mode $v$ remains to be optimized.

Figure~\ref{fig:local_sensitivity}(b) shows how the sensitivity to each control evolves with the coupling. Perturbations that act on the bright mode ($n_+$), or that mix the bright and dark modes (detuning), lose their sensitivity at strong coupling, since the occupation of the bright mode becomes pinned and the bright--dark coherence vanishes. The dark-mode occupation $n_-=\frac12(N_S-d_1^\dagger d_2-d_2^\dagger d_1)$, a combination of common shift and tunneling, instead retains a sensitivity independent of the coupling, and the optimum over all single-mode occupations converges to it at strong coupling. It yields
\begin{equation}
    \tau\geq\tau_{\rm Pl}\sqrt{f_\beta(\bar\epsilon)\big[1-f_\beta(\bar\epsilon)\big]},
    \label{eq:dqd_dark_bound}
\end{equation}
a Planckian bound independent of the coupling. The normalized common shift gives half of this value at strong coupling. Both perturbations shift the dark-mode energy, which is the information retained by the equilibrium state, but $N_S/2=(n_++n_-)/2$ spends half of its norm on the pinned bright mode (Appendix~\ref{app:fermionic}).

\paragraph*{Optimal observable.} Figure~\ref{fig:local_sensitivity}(c) shows which observables recover this sensitivity through their mean response, Eq.~\eqref{eq:qfi_variational}. At weak coupling the dot charges suffice: for the common shift, the total charge recovers the full QFI. At strong coupling, the variance of the total charge contains the large fluctuations of the pinned bright mode, which carry no response, and only the dark occupation recovers the full QFI, for both perturbations. For $\kappa=n_-$, measuring $n_-$ itself is optimal at every coupling, so that Eq.~\eqref{eq:self_response_bound} is exact and, at strong coupling, the optimal accessible perturbation and its optimal readout coincide, $\kappa^*\simeq A^*\simeq n_-$, as anticipated in Sec.~\ref{subsec:optimal_perturbations}. Since $[n_-,V]=0$ and $\beta\Delta\ll1$, the fluctuations of $n_-$ evolve slowly compared with $\hbar\beta$, so that $r_{n_-}\simeq1$ and the bound is Planckian, an instance of fluctuations that are slow but not frozen. Measuring $n_-$ requires the interdot coherence $\mathrm{Re}\langle d_1^\dagger d_2\rangle$ in addition to the dot charges.

The near conservation of $n_-$ that makes the bound Planckian is also what makes relaxation slow: the most sensitive direction acts on the mode that is most weakly coupled to the reservoir.  We therefore perform the finite-error test along the common shift, $\hat\kappa=N_S/2$, a gate voltage applied to both dots. Since $N_S=\sum_nd_n^\dagger d_n$, it adds a multiple of the identity to $h$ and $\Gamma$, shifting both levels without changing the normal modes or their widths.

\textit{Finite-error test}. Figure~\ref{fig:finite_error}(b) shows the common-shift family, with all members prepared in $(n_1,n_2)=(1,0)$. The thermalization time is nonmonotonic in the coupling. At weak coupling, it decreases as $\hbar/g^2$, as each dot relaxes into the reservoir. It is shortest near $g^2\simeq|\Delta|$, where the two eigenvalues of $\Gamma$ coalesce, and it then grows linearly with $g^2$, as the dark mode relaxes on the time scale $2\hbar g^2/\Delta^2$. Since the common shift leaves the decay rates unchanged, this coupling dependence is that of $\bar H_S$ for every accuracy; the accuracy only sets how many relaxation times are required. The bound, instead, approaches a coupling-independent constant at every accuracy. In contrast to the RLM, the fixed task therefore never becomes trivial: the dark mode keeps the targets distinguishable at strong coupling. Strong coupling thus neither removes the Planckian constraint nor makes thermalization fast. The bound remains far below $\tau_\varepsilon$, because it constrains only the acquisition of the surviving equilibrium information, whereas the slow dark mode delays the dynamics.

Along the dark-mode direction, $\hat\kappa=n_-$, the bound is larger at every coupling and approaches twice its common-shift value at strong coupling, but thermalization is slower by orders of magnitude, except for $g^2\gtrsim\delta$, where both families relax at the same rate (Appendix~\ref{app:dark_family}, Fig.~\ref{fig:dark_comparison}). The reason is that shifting $n_-$ detunes the dark level from the bright mode, through which it relaxes. In this model, both effects stem from the same protected mode, so that the optimal perturbation defines a harder task rather than a tighter bound.

\subsection{Benchmark against open-system speed limits}
\label{subsec:speed_limits}

In the fermionic models the exact dynamics is available, so $\tau_\varepsilon\geq\tau_{\rm SS}$ and $\tau_\varepsilon\geq\tau_{\rm Pl}\chi_\delta$ can be compared for the same task: the same family, preparation, and accuracy as in Fig.~\ref{fig:finite_error}. Here $\tau_{\rm SS}$ is the maximum in Eq.~\eqref{eq:ss_bound} over the family $H_S(x)$, $|x|\leq\delta$, and $\mathcal D_x\equiv\mathcal D(H_S(x))$. This comparison favors the speed limit, which uses the energy fluctuations along the exact trajectories, whereas our bound uses only the equilibrium targets. Its outcome is nevertheless dictated by the bandwidth. The reservoir acts on the levels through a single collective mode, and in the factorized initial state the interaction contributes $\operatorname{Var}(V)=Ng^2\Lambda/4$ to $\delta E_S^2$ for every preparation, while $\|H_S+V\|\simeq g\sqrt{2N\Lambda}$ (Appendix~\ref{app:fermionic}). Once this contribution dominates, both forms of the speed limit decrease as $(g^2\Lambda)^{-1/2}$. They depend on an ultraviolet scale that leaves the MFG targets and the dynamics at $t\gg\hbar/\Lambda$, and hence $\tau_\varepsilon$ and $\chi_\delta$, essentially unchanged, and they vanish at strong coupling irrespective of $\tau_\varepsilon$. In the DQD, in particular, they vanish where $\tau_\varepsilon$ grows with the coupling and $\chi_\delta$ remains finite. Since any finite value of $\tau_{\rm SS}$ reflects the chosen $\Lambda$, a comparison along the coupling at a single bandwidth would be misleading; Appendix~\ref{app:fermionic} shows it for several bandwidths (Fig.~\ref{fig:speedlimit_sensitivity}).

Together, the two models delimit the scope of the Planckian constraint at strong coupling. The constraint disappears when the reduced equilibrium state forgets the Hamiltonian, as in the RLM, and then for every admissible perturbation and measurement. When the system--bath structure protects part of that information, as the dark mode does in the DQD, a Planckian bound independent of the coupling survives. The optimal perturbation and the optimal measurement can then be identified, and the bound remains finite where speed limits based on the interaction energy vanish.

\section{Spin--boson model: reorganization of equilibrium information}
\label{sec:spin_boson}

The fermionic models are quadratic, so that their MFG states and their dynamics follow in closed form. The bound does not rely on this structure. It requires only the MFG targets and, for a qubit, only the MFG Bloch vector and its static response matrix (see case~(iii) of Sec.~\ref{subsec:optimal_perturbations}). We now apply it to the spin--boson model~\cite{CresserAnders2021MFG, PhysRevA.106.042209}, whose equilibrium state and dynamics have no analytical expression at arbitrary coupling. Both are accessible numerically, but only the equilibrium state enters the bound; the dynamics serves only to test it. The model shows how strong coupling reorganizes the Planckian constraint: as the coupling grows, only the part of a Hamiltonian change along the axis through which the bath couples to the spin eventually remains visible in equilibrium.

We consider a spin coupled through $\sigma_z$ to a bath of harmonic oscillators,
\begin{align}
    H_S(\theta)&=\frac{\Delta}{2}\left(\sin\theta\,\sigma_x+\cos\theta\,\sigma_z\right),
    \qquad
    H_B=\sum_k\Omega_k\,b_k^\dagger b_k,
    \nonumber\\
    V&=g\,\sigma_z\otimes B,
    \qquad
    B=\sum_k\left(t_k\,b_k^\dagger+t_k^*\,b_k\right),
    \label{eq:sb_hamiltonian}
\end{align}
where  the bath enters only through its spectral density $\mathcal J(\Omega)=\pi\sum_k|t_k|^2\delta(\Omega-\Omega_k)$. The coupling selects the pointer basis $\{|{\pm z}\rangle\}$, and $\theta$ is the angle between $H_S$ and this axis. At $\theta=\pi/2$, Eq.~\eqref{eq:sb_hamiltonian} is the unbiased spin--boson model with tunneling energy $\Delta$; otherwise, $\Delta\cos\theta$ acts as a bias. At $\theta=0$, $\sigma_z$ commutes with $H_{\rm tot}$ and the pointer populations never relax, so that Definition~\ref{def:thermalization} cannot be met (Sec.~\ref{subsec:mfg_qfi}); we take $0<\theta<\pi$.

The MFG state is known analytically only in two limits. At weak coupling, it deviates from the bare Gibbs state at second order in $g$~\cite{CresserAnders2021MFG,Glatthard2023EnergyMeasurements},
\begin{align}
    \tilde\omega&=\omega(\beta,H_S)\big[\mathbb{I}+g^2X\big]+O(g^4),
    \nonumber\\
    X&=\int_0^\beta\!d\lambda\int_0^\lambda\!d\lambda'\,\Phi_B(\lambda-\lambda')
    \nonumber\\
    &\quad\times\big[\sigma_z(\lambda)\sigma_z(\lambda')-\langle\sigma_z(\lambda)\sigma_z(\lambda')\rangle\big],
    \label{eq:sb_wbf}
\end{align}
where $\sigma_z(\lambda)=e^{\lambda H_S}\sigma_ze^{-\lambda H_S}$, the average is taken in $\omega(\beta,H_S)$, and $\Phi_B(\lambda)=\pi^{-1}\int_0^\infty d\Omega\,\mathcal J(\Omega)\cosh[\Omega(\beta/2-\lambda)]/\sinh(\beta\Omega/2)$ is the imaginary-time correlation function of $B$ in $\omega(\beta,H_B)$. The expansion holds as long as $g^2X$ remains small. At ultrastrong coupling, the bath dephases the spin in its pointer basis, and the MFG state becomes the Gibbs state of the pinched Hamiltonian~\cite{CresserAnders2021MFG,PhysRevA.106.042209},
\begin{equation}
    \tilde\omega\to\frac{e^{-\beta\mathcal Z[H_S]}}{\Tr\,e^{-\beta\mathcal Z[H_S]}},
    \qquad
    \mathcal Z[H_S]=\frac{\Delta\cos\theta}{2}\,\sigma_z,
    \label{eq:sb_usc}
\end{equation}
where $\mathcal Z[Y]=\sum_{s=\pm}P_sYP_s$, with $P_\pm=|{\pm z}\rangle\langle{\pm z}|$, removes the coherences in the pointer basis. Only the bias survives.

The thermalization task rotates $H_S=\frac12\bm h\cdot\bm\sigma$, whose field $\bm h=\Delta(\sin\theta,0,\cos\theta)$ makes an angle $\theta$ with the coupling axis. Locally, a rotation by $\delta\theta$ gives $H_S(\theta+\delta\theta)=H_S(\theta)+\delta\,\kappa_\theta$, with $\delta=\Delta\,\delta\theta$ and
\begin{equation}
    \kappa_\theta={\frac{1}{\Delta}}\,\partial_\theta H_S=\frac{\,{1}}{2}\,\bm n_\theta\cdot\bm\sigma,
    \qquad
    \bm n_\theta=(\cos\theta,0,-\sin\theta),
    \label{eq:sb_rotation}
\end{equation}
and $\|\kappa_\theta\|=\,{1}$; the rotation tilts $\bm h$ within the $xz$ plane, $\bm n_\theta\perp\bm h$. For a perturbation $\kappa=\frac12\bm n\cdot\bm\sigma$, case~(iii) gives $\mathcal F^{\rm th}_\kappa=\bm n^\top K\bm n$, with $K=J(\mathbb{I}+\bm b\bm b^\top/(1-|\bm b|^2))J$ the matrix in Eq.~\eqref{eq:qubit_chi}. The rotation thus yields $\tau\geq\tau_{\rm Pl}\sqrt{\bm n_\theta^\top K\bm n_\theta}/\beta$, and the optimal perturbation yields $\tau\geq\tau_{\rm Pl}\widetilde\chi$, with $\widetilde\chi=\sqrt{\lambda_{\max}(K)}/\beta$.

At the symmetric point $\theta=\pi/2$, the rotation is a bias, $\kappa_{\pi/2}=-\frac{\,{1}}{2}\sigma_z$, and $H_{\rm tot}$ is invariant under a $\pi$ rotation of the spin about $\hat{\bm x}$ combined with the bath parity $b_k\to-b_k$. This symmetry flips $\sigma_z$, so that $\langle\sigma_z\rangle_{\tilde\omega}=0$ and $\bm b$ points along $\hat{\bm x}$. This is the situation of case~(iii) in which $\bm b$ is an eigenvector of $J$, so that Eq.~\eqref{eq:qubit_channels} applies at every coupling. The bias is perpendicular to $\bm b$, its SLD is proportional to $\sigma_z$, and measuring the pointer populations is therefore optimal at every coupling. The bound reduces to a single isothermal susceptibility, defined in Eq.~\eqref{eq:static_susceptibility},
\begin{equation}
    \tau\geq\frac{\hbar}{2}\,\xi_{\sigma_z\sigma_z}
    =-\hbar\,\frac{\partial\langle\sigma_z\rangle_{\tilde\omega}}{\partial h_z}\bigg|_{h_z=0},
    \label{eq:sb_symmetric}
\end{equation}
where $h_z$ is the component of $\bm h$ along the coupling axis. Since $\operatorname{Var}_{\tilde\omega}(\sigma_z)=1$, Eq.~\eqref{eq:qfi_hierarchy} gives $\xi_{\sigma_z\sigma_z}\leq\beta$, so that Eq.~\eqref{eq:sb_symmetric} can approach, but not reach, the ceiling $\tau_{\rm Pl}/2$. The rotation and its optimal readout are thus both along $\sigma_z$ at every coupling, which makes the symmetric point the natural place for the finite-error test below.

\begin{figure*}[t]
    \centering
    \includegraphics[width=\textwidth]{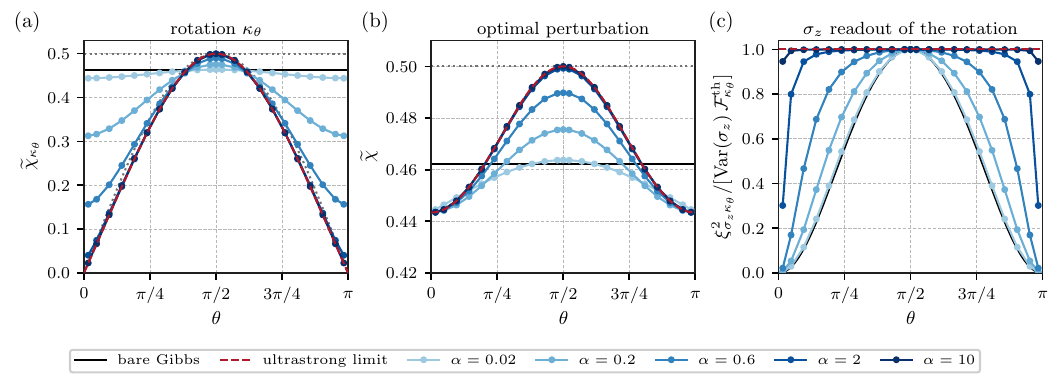}
    \caption{\textbf{Equilibrium constraint of the spin--boson model at all couplings.}
    Locally exact bounds, in units of $\tau_{\rm Pl}$, versus the angle $\theta$ between $H_S$ and the coupling axis, from MFG states computed with imaginary-time TEMPO at increasing Ohmic coupling $\alpha$ (blue), together with the bare Gibbs state (black) and the ultrastrong limit, Eq.~\eqref{eq:sb_usc_bound} (red dashed). Dotted: the ceiling $1/2$ and, in (a), $|\sin\theta|/2$.
    (a)~Rotation $\kappa_\theta$, $\widetilde\chi_{\kappa_\theta}=\sqrt{\bm n_\theta^\top K\bm n_\theta}/\beta$.
    (b)~Optimal perturbation, $\widetilde\chi$.
    (c)~Fraction of $\mathcal F^{\rm th}_{\kappa_\theta}$ recovered by measuring $\sigma_z$, $\xi_{\sigma_z\kappa_\theta}^2/[\operatorname{Var}_{\tilde\omega}(\sigma_z)\,\mathcal F^{\rm th}_{\kappa_\theta}]$, which follows from Eq.~\eqref{eq:qfi_variational}.
    Here $\beta\Delta=1$ and $\Omega_c=10\Delta$.}
    \label{fig:sb_equilibrium}
\end{figure*}

\paragraph*{Optimal perturbation and observable.}
At zero coupling, $\tilde\omega$ is the Gibbs state of $H_S$, which is invariant under rotations about $\bm h$, so that $J$ responds equally to all perturbations perpendicular to $\bm h$. With $x_q=\beta\Delta/2$, one has $\bm b=-\tanh(x_q)\,\hat{\bm h}$, and the rotation, being perpendicular to $\bm h$, gives $\mathcal F^{\rm th}_{\kappa_\theta}=\tanh^2x_q{/\Delta^2}$ and $\tau\geq\tau_{\rm Pl}\tanh(x_q)/(2x_q)$ at every $\theta$ (black lines in Fig.~\ref{fig:sb_equilibrium}). This is the transverse value of Eq.~\eqref{eq:qubit_channels}: the rotation is an optimal perturbation. Its SLD is proportional to $\bm n_\theta\cdot\bm\sigma$, a coherence in the eigenbasis of $H_S$, so that measuring $\sigma_z$ recovers only part of the information, as can be seen in Fig.~\ref{fig:sb_equilibrium}(c).

At weak but finite coupling, Eq.~\eqref{eq:sb_wbf} gives $\bm b$ and $J$, and thus both bounds, to second order in $g$. The coupling breaks this equivalence at once: it suppresses the response to perturbations perpendicular to the coupling axis and enhances the response along it. The bound for the rotation therefore decreases near alignment, where $\bm n_\theta$ is nearly perpendicular to $\hat{\bm z}$, and increases around $\theta=\pi/2$, where the rotation is a bias, as shown by the weakest coupling in Fig.~\ref{fig:sb_equilibrium}(a). In this regime the bound follows analytically, whereas a consistent description of the dynamics is demanding: a master equation whose stationary state is correct to second order in $g$ must itself be of fourth order~\cite{Fleming2011PerturbativeME}, and Lindblad approximations generically miss these corrections~\cite{Tupkary2022LindbladLimitations}.

At ultrastrong coupling, Eq.~\eqref{eq:sb_usc} retains only the bias $\Delta\cos\theta$. A perturbation $\frac12\bm n\cdot\bm\sigma$ survives the pinching only through $n_z$, so that $J\to-\frac\beta2{\rm sech}^2(x_\theta)\,\hat{\bm z}\hat{\bm z}^\top$, with $x_\theta=\beta\Delta\cos\theta/2$. Only the longitudinal channel of Eq.~\eqref{eq:qubit_channels} remains, and
\begin{equation}
    \widetilde\chi\to\frac12\,{\rm sech}\,x_\theta,
    \qquad
    \mathcal F^{\rm th}_{\kappa_\theta}\to\frac{\beta^2}{4}\sin^2\theta\,{\rm sech}^2x_\theta.
    \label{eq:sb_usc_bound}
\end{equation}
The optimal perturbation is the bias $\sigma_z/2$, and its optimal readout is $\sigma_z$, i.e., the pointer populations. The rotation changes the bias only at a rate $\propto\sin\theta$ and falls short of the optimum by the factor $|\sin\theta|$: near alignment, it tilts $H_S$ perpendicular to the coupling axis, a change that the bath erases. Since $\tanh x\leq x\leq\sinh x$, the optimal bound exceeds its zero-coupling value near $\theta=\pi/2$, where it approaches the ceiling $\tau_{\rm Pl}/2$, and falls below it near alignment.

Between these limits, neither the MFG state nor the dynamics has an analytical expression. We compute the MFG state at arbitrary coupling with the imaginary-time version~\cite{Chiu2022MFGNumerics} of the time-evolving matrix product operator (TEMPO) algorithm~\cite{Strathearn2018TEMPO}, a numerically exact method for Gaussian baths, as implemented in OQuPy~\cite{Fux2024OQuPy}; $\bm b$ and $J$ follow from it.
We take an Ohmic bath, $g^2\mathcal J(\Omega)=\frac\pi2\alpha\,\Omega\,e^{-\Omega/\Omega_c}$, where $\alpha$ is the dimensionless coupling of Ref.~\cite{Leggett1987SpinBoson}. Figure~\ref{fig:sb_equilibrium} shows that the numerics connect the two limits continuously. As the coupling grows, Fig.~\ref{fig:sb_equilibrium}(a) shows the flat profile of the rotation bound deforming into the $|\sin\theta|$ profile of Eq.~\eqref{eq:sb_usc_bound}, and Fig.~\ref{fig:sb_equilibrium}(c) shows that measuring $\sigma_z$ recovers an increasing fraction of the information, all of it in the ultrastrong limit: the information carried by a rotation moves from the coherences to the pointer populations. In contrast, as can be seen in Fig.~\ref{fig:sb_equilibrium}(b), the optimal bound remains Planckian at every angle and coupling. The coupling only redistributes it between orientations, while the optimal perturbation migrates from the transverse channel to the bias $\sigma_z/2$.

\begin{figure}[t]
    \centering
    \includegraphics[width=\columnwidth]{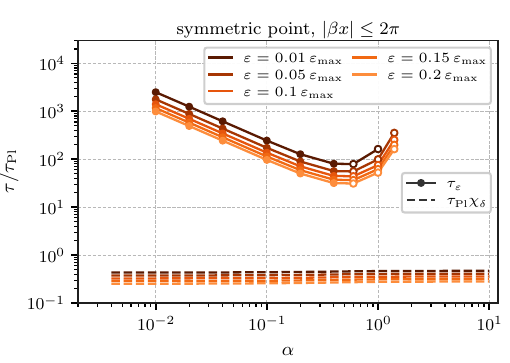}
    \caption{\textbf{Finite-error test at the symmetric point.}
    Thermalization time $\tau_\varepsilon$ from real-time TEMPO (solid) and bound $\tau_{\rm Pl}\chi_\delta$ from the MFG targets (dashed) versus the coupling, for the family $H_S(x)=\frac\Delta2\sigma_x+\frac x2\sigma_z$ with $|\beta x|\leq2\pi$, the preparation $\rho_S(0)=\mathbb{I}/2$, and the accuracies of Sec.~\ref{subsec:fermionic_model}. Open markers carry the residual uncertainty of the real-time numerics, which are not converged at larger couplings. Parameters as in Fig.~\ref{fig:sb_equilibrium}.}
    \label{fig:sb_task}
\end{figure}

\paragraph*{Finite-error test.}  We perform this test at the symmetric point and extend the bias direction as in Eq.~\eqref{eq:hamiltonian_family}: $H_S(x)=\frac\Delta2\sigma_x+\frac x2\sigma_z$, i.e., $x=h_z$, with the same range $\delta$ and accuracies $\varepsilon$ as in Sec.~\ref{subsec:fermionic_model} and the preparation $\rho_S(0)=\mathbb{I}/2$. We obtain $\tau_\varepsilon$ from real-time TEMPO~\cite{Strathearn2018TEMPO,Fux2024OQuPy} combined with transfer tensors~\cite{CerrilloCao2014TTM}. Figure~\ref{fig:sb_task} shows that the bound, computed from the MFG targets alone, is finite at every coupling and lies below $\tau_\varepsilon$ wherever the latter is available. It hardly depends on the coupling, for a reason that can be stated exactly. The optimal pair is symmetric, $(-x^*,x^*)$, and the symmetry of the symmetric point gives its two targets the same coherence $b_x$ and opposite pointer polarizations $\pm b_z$. Their Bures angle is then $\arcsin|b_z|$. The coherence, which the coupling destroys, drops out, and for $k_BT\gtrsim\Delta$ the pointer polarization of the biased members changes little between the bare and the pinched states. Rotating $H_S$ instead of biasing it gives nearly the same bound at the strictest accuracies.

The thermalization time behaves as in the DQD. It decreases as $1/\alpha$ at weak coupling and grows at strong coupling, because the equilibrium information is carried by $\sigma_z$, which commutes with $V$, as the dark occupation does in the DQD. The pointer populations then relax only through the tunneling, which the bath dresses and suppresses~\cite{Leggett1987SpinBoson,Weiss2012QuantumDissipativeSystems}.
The bound does not capture this slowdown, since it constrains only the acquisition of the equilibrium information. Speed limits provide no comparison here. The Shiraishi--Saito bound requires the energy fluctuations of $H_S+V$ in the state of the compound, which methods for the reduced dynamics do not provide, and its static form in Eq.~\eqref{eq:ss_chain} vanishes because $\|V\|$ is unbounded. The trajectory form requires the reduced dynamics up to $\tau_\varepsilon$, and hence $\tau_\varepsilon$ itself.

\begin{figure*}[t]
    \centering
    \includegraphics[width=\textwidth]{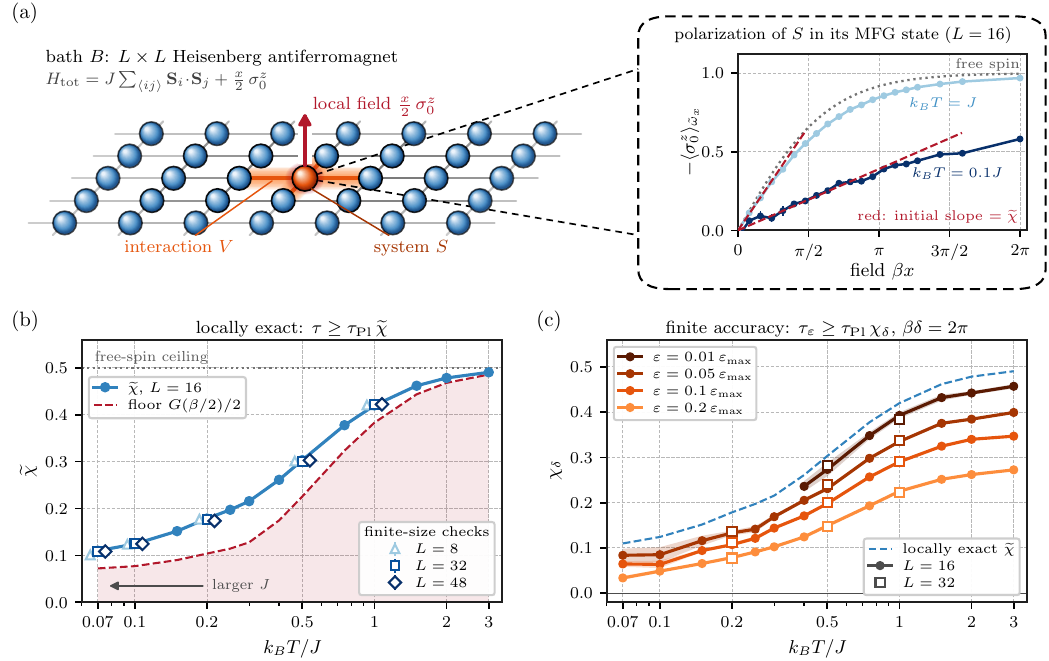}
    \caption{\textbf{Planckian bound for a spin in a two-dimensional antiferromagnet, from equilibrium data alone.}
    (a)~Setup of Eq.~\eqref{eq:heisenberg}: the system $S$ is one spin of an $L\times L$ Heisenberg antiferromagnet, coupled to its four neighbors through $V$ and perturbed by a local field $\frac{x}{2}\sigma^z_0$. Inset: polarization of $S$ in its MFG state versus the field, for $L=16$. Its initial slope (red dashed) is the locally exact bound $\widetilde\chi$, Eq.~\eqref{eq:heisenberg_chi}, and the free spin (dotted) gives the ceiling $1/2$.
    (b)~Locally exact bound $\widetilde\chi$ versus temperature; at fixed temperature, lower $k_BT/J$ corresponds to stronger coupling. Line: $L=16$. Open symbols: $L=8$, $32$, and $48$, slightly offset horizontally for visibility. Dashed red line: the floor $G(\beta/2)/2$ of Eq.~\eqref{eq:heisenberg_floor}, set by the slow fluctuations of $\sigma^z_0$; the shaded region below it is excluded.
    (c)~Finite-accuracy bound $\chi_\delta$ for $|\beta x|\leq2\pi$, evaluated on pairs of simulated fields, for $L=16$ (lines) and $L=32$ (open squares), with $\widetilde\chi$ dashed. Points whose statistical uncertainty exceeds $20\%$ of their value are not shown.
    Bands: one standard deviation.}
    \label{fig:heisenberg}
\end{figure*}

\section{Two-dimensional antiferromagnet: the bound from equilibrium data alone}
\label{sec:heisenberg}

We now evaluate the bound for a system whose equilibrium state is accessible but whose long-time dynamics is not: a spin in a two-dimensional quantum antiferromagnet. Its MFG state and susceptibilities follow from quantum Monte Carlo without a sign problem, whereas its real-time evolution is limited to short times in tensor-network methods by the growth of entanglement~\cite{Schuch2008EntropyScaling,Czarnik2019iPEPSTimeEvolution,Hubig2019iPEPSDisorder} and requires an ill-posed analytic continuation in Monte Carlo~\cite{Jarrell1996AnalyticContinuation}.

We consider the spin-$\frac12$ Heisenberg antiferromagnet on an $L\times L$ square lattice with periodic boundary conditions, in which the spin at site $0$ is the system, as sketched in Fig.~\ref{fig:heisenberg}(a):
\begin{equation}
    H_{\rm tot}=\underbrace{\tfrac{x}{2}\,\sigma^z_0}_{H_S(x)}
    +\underbrace{J\,\mathbf S_0\cdot\sum_{j\in\partial0}\mathbf S_j}_{V}
    +\underbrace{J\sum_{\langle ij\rangle\not\ni0}\mathbf S_i\cdot\mathbf S_j}_{H_B},
    \label{eq:heisenberg}
\end{equation}
with $J>0$ and $\partial0$ the four neighbors of site $0$. Since $V$ and $H_B$ have the same strength, a stronger coupling at fixed temperature amounts to a lower $k_BT/J$. At $x=0$, spin-rotation symmetry gives exactly $\tilde\omega=\mathbb{I}/2$ and an isotropic static response matrix in case~(iii) of Sec.~\ref{subsec:optimal_perturbations}. Equation~\eqref{eq:qubit_channels} then gives the optimal locally exact bound over all perturbations of the spin,
\begin{equation}
    \widetilde\chi=\frac{\xi_{\sigma\sigma}}{2\beta}
    =-\frac{\partial\langle\sigma^z_0\rangle_{\tilde\omega_x}}{\partial(\beta x)}\bigg|_{x=0},
    \label{eq:heisenberg_chi}
\end{equation}
with $\xi_{\sigma\sigma}$ the static susceptibility of $\sigma^z_0$. The bound is thus the initial slope of the polarization of $S$ versus $\beta x$, shown in the inset of Fig.~\ref{fig:heisenberg}(a); a free spin, with $\langle\sigma^z_0\rangle=-\tanh(\beta x/2)$, attains the ceiling $1/2$. We compute these quantities with the stochastic series expansion (SSE) quantum Monte Carlo method~\cite{SandvikKurkijarvi1991,Sandvik1999OperatorLoop,Sandvik2010ComputationalSpin,Sandvik2019SSE,Merali2024SSERydberg} with directed-loop updates~\cite{SyljuasenSandvik2002DirectedLoops}. It samples $\Tr e^{-\beta H_{\rm tot}}$ without Trotter or truncation errors and, after a sublattice rotation~\cite{Marshall1955Antiferromagnetism}, without a sign problem for any field. The same samples give the populations of $S$, i.e., its MFG state, and, through the standard SSE estimators~\cite{SandvikKurkijarvi1991, Sandvik1992Handscomb}, the susceptibility $\xi_{\sigma\sigma}$ and the imaginary-time correlation $G(\lambda)$ used below.

All data in Fig.~\ref{fig:heisenberg} are equilibrium estimators, exact up to statistical errors, and the code was validated against exact diagonalization of small clusters. We use $L=16$ and check the dependence on the lattice size with $L=8$, $32$, and $48$ at five temperatures, mostly at low temperature, where the antiferromagnetic correlation length grows rapidly~\cite{Chakravarty1989TwoDimensionalHeisenberg,Kim1998SquareLatticeHeisenberg}. For the finite-accuracy bound, with the same range $\delta$ and accuracies $\varepsilon$ as in Sec.~\ref{subsec:fermionic_model}, we evaluate $\chi_\delta$ only on pairs of simulated fields. Restricting the pairs can only lower the maximum, so the result is a valid bound.

Figure~\ref{fig:heisenberg}(b) shows the locally exact bound. At high temperature, the spin is nearly free and $\widetilde\chi$ approaches the ceiling $1/2$. As the coupling grows, the bound decreases but, unlike in the resonant-level model, where $\widetilde\chi\to4/(\pi\beta g^2)$, it remains above a finite floor. Lattices with $L=16$, $32$, and $48$ agree within $3\%$, so the bound has converged in the lattice size; only the smallest one, $L=8$, falls below them at the lowest temperature.

The bound survives because part of the fluctuations of $\sigma^z_0$ is slower than the Planckian time. Imaginary-time correlations isolate this part. For $\kappa=\sigma^z_0/2$, Eq.~\eqref{eq:r_floor} gives $r_\kappa\geq G(\beta/2)$, where $G(\lambda)=\langle\sigma^z_0(\lambda)\,\sigma^z_0\rangle$ at $x=0$. As Eq.~\eqref{eq:self_response_bound} is exact here,
\begin{equation}
    \widetilde\chi=\frac{r_\kappa}{2}\geq\frac{G(\beta/2)}{2}.
    \label{eq:heisenberg_floor}
\end{equation}
This floor requires a single equilibrium correlator. As can be seen in Fig.~\ref{fig:heisenberg}(b), the floor remains finite down to the lowest temperature simulated, $k_BT=0.07J$, i.e., up to $J\simeq14\,k_BT$. The gap between $\widetilde\chi$ and the floor, due to the fast fluctuations, instead closes roughly linearly in $k_BT/J$, as expected for fluctuations at frequencies of order $J/\hbar$. The bound therefore remains Planckian at strong coupling, $\tau\gtrsim0.1\,\hbar/k_BT$. Unlike in the DQD and the spin--boson model, $\sigma^z_0$ does not commute with the interaction. Its slow fluctuations are a collective property of the bath, whose antiferromagnetic correlations extend over large regions at low temperature~\cite{Chakravarty1989TwoDimensionalHeisenberg,Manousakis1991HeisenbergReview}: the spin follows a locally ordered background whose orientation changes slowly.

As shown in Fig.~\ref{fig:heisenberg}(c), the finite-accuracy bound follows $\widetilde\chi$, reduced by a factor that grows with $\varepsilon$, and remains finite at all temperatures, with the same values for $L=16$ and $L=32$. Every quantity in Fig.~\ref{fig:heisenberg} is an equilibrium property of the compound, and the resulting constraint holds for any dynamics that thermalizes the spin, which we never had to compute.

\section{Conclusions and outlook}
\label{sec:conclusion}

Can strong system–bath coupling make thermalization faster than the Planckian time~$\tau_{\rm Pl}\equiv  \hbar/k_B T$? Our results show that equilibrium properties alone can rule this out. Our bound falls below the Planckian scale only if the mean-force Gibbs (MFG) state becomes insensitive to the system Hamiltonian, or encodes it only in observables whose fluctuations decorrelate faster than~$\tau_{\rm Pl}$. Otherwise, it remains Planckian at strong coupling, even where speed limits become trivial. When the full MFG family or its QFI is not accessible, the static susceptibility and variance of a single perturbation of $H_S$ still yield a valid bound on the thermalization time. Both results follow from requiring thermalization to be robust, i.e., to work for a whole family of system Hamiltonians, so that it must acquire information about which one is present~\cite{Abiuso2025PlanckianBound}. Technically, the bound follows from equilibrium correlations of the global state, without the mean-force Hamiltonian, and the strongest bound is attained by perturbing $H_S$ with a projector, i.e., by shifting the energy of a subspace relative to its complement.

The models we studied illustrate this criterion. In the resonant-level model, strong coupling makes the targets indistinguishable at finite accuracy, and the occupation decorrelate on times much shorter than $\tau_{\rm Pl}$, so that the locally exact bound loses its Planckian character. It becomes comparable to the occupation relaxation time $\hbar/g^2$, and the exact thermalization time follows the same trend (Fig.~\ref{fig:finite_error}). In the double quantum dot and the spin--boson model, the information is carried by observables (the dark occupation and the pointer polarization) that commute with the interaction and relax only through $H_S$, so their fluctuations remain slow and the bound stays Planckian at every coupling. The two-dimensional antiferromagnet shows that commuting with the interaction is not necessary: the observable $\sigma_0^z$ does not commute with $V$, yet it retains slow fluctuations inherited from the collective correlations of the bath. The bound there decreases with the coupling but remains of order $\tau_{\rm Pl}$ over the range explored (Fig.~\ref{fig:heisenberg}), in contrast to the resonant-level model.

A central feature of the bound is its generality. Apart from the equilibrium targets, it uses no information about the interaction, the bath, or the initial state, so it holds for any dynamics that thermalizes the family of Hamiltonians. This distinguishes it from open-system speed limits~\cite{Shiraishi2021SpeedLimit}, which bound the time needed to prepare a given target from a given initial state, and can become trivial as the interaction-energy fluctuations grow (Sec.~\ref{subsec:speed_limits_general}). The two bounds constrain different costs, preparation versus information, and are complementary (Table~\ref{tab:bound_nature}). Since the bound requires only equilibrium information, it can also be evaluated when the dynamics is out of reach. The two-dimensional quantum magnet of Sec.~\ref{sec:heisenberg} illustrates this. Its thermalization dynamics is expected to be out of reach of tensor-network methods, which are confined to short times by the growth of entanglement~\mbox{\cite{Schuch2008EntropyScaling,Czarnik2019iPEPSTimeEvolution,Hubig2019iPEPSDisorder}}, and of quantum Monte Carlo, which requires an ill-posed analytic continuation~\mbox{\cite{Jarrell1996AnalyticContinuation}}, whereas its MFG state and susceptibilities follow from sign-problem-free quantum Monte Carlo, so that our bound constrains any such dynamics without computing it. The price of this generality is that the bound need not be tight: it constrains the acquisition of information, not the full relaxation. In the double quantum dot, for instance, the thermalization time grows with the coupling while the bound does not.

Several directions follow. The bound can be refined by including information about the process, for instance by combining the equilibrium sensitivity with metrological bounds on noisy dynamics~\cite{Grecki2026Time}. Experimentally, the bound can be evaluated from the equilibrium response of a system observable to a control parameter, or from its fluctuation spectrum, as measured by noise spectroscopy or neutron scattering~\cite{Crooker2004SpinNoise,Scheie2021NeutronEntanglement}. The same spectra have been used to witness multipartite entanglement through the QFI of unitarily evolved thermal states~\cite{Hauke2016DynamicSusceptibility,Scheie2021NeutronEntanglement}. Such measurements would give a lower bound on the thermalization time of a complex system without modeling its environment. 

Planckian time scales have also been recently identified, in thermal states, through different routes: static properties of the Gibbs state lower bound scrambling times in SYK and random spin models~\cite{Ma2026StaticBounds}; charge imbalances between subsystems of SYK-type models relax at a Planckian rate~\cite{Kim2026PlanckianCharge}; and, for general local many-body systems, the onset time of hydrodynamics obeys a Planckian lower bound~\cite{Qi2026PlanckianLocal,delacretaz2025bound}. These works rely on analytic properties of thermal correlators, on hydrodynamic fluctuations, or on large-$N$ conformal methods, and target specific time scales, such as scrambling or the onset of hydrodynamics. In contrast, our approach uses quantum metrology and lower-bounds the time to reach equilibrium under any dynamics that thermalizes robustly, without assuming locality. Bringing these perspectives closer could shed further light on the origin of Planckian time scales in scrambling, transport and equilibration~\cite{Hartnoll2022PlanckianDissipation}.

\section*{Acknowledgments}
The authors thank Paolo Abiuso and Alberto Rolandi for fruitful discussions.  Anthropic's Claude (Opus~5.5) was used as a coding assistant for the quantum Monte Carlo and TEMPO computations and to edit the text; all code was reviewed and validated by the authors, who take full responsibility for this work. 
J.~J.-R.  and M.P.-L. acknowledges financial support from the Spanish Agencia Estatal de Investigación (AEI) through the 
project a Project PID2024-162153NB- I00 funded by MICIU/AEI/10.13039/501100011033 and by ERDF, EU. M.P.-L. also acknowledges  funding from  the ATRAE Program (Grant ATR2024-154621) funded by MICIU/AEI/10.13039/501100011033. Additional support was provided by grant PID2022-141283NB-I00 (MICIU/AEI/10.13039/501100011033). J.C. acknowledges support from ICREA Academia.

\section*{Data availability}
All code for the simulations is available in Ref.~\cite{git}.

\bibliography{references}
\onecolumngrid

\appendix
\renewcommand{\thesection}{\Alph{section}}
\renewcommand{\thesubsection}{\roman{subsection}}
\renewcommand{\theequation}{\thesection.\arabic{equation}}
\renewcommand{\thefigure}{S\arabic{figure}}
\setcounter{figure}{0}

\section{Equilibrium sensitivity of mean-force Gibbs states}
\label{app:mfg_sensitivity}

This appendix proves the identities of Sec.~\ref{sec:mfg_sensitivity}. We keep the notation of Sec.~\ref{subsec:mfg_response} and write $\langle X\rangle=\Tr[\omega_{\rm tot}X]$, which equals $\Tr[\tilde\omega X]$ for system operators. We take $\omega_{\rm tot}$ to be trace class; for a macroscopic bath, the identities hold for the limits of the equilibrium correlations involved. Appendix~\ref{app:mfg_response} derives the static susceptibility $\xi_{A\kappa}$, App.~\ref{app:spectral} writes it as a filtered integral of the equilibrium fluctuation spectrum, App.~\ref{app:proof_prop} proves Proposition~\ref{prop:mfg_qfi}, and App.~\ref{app:qubit} works out the qubit case.

Both Apps.~\ref{app:spectral} and~\ref{app:proof_prop} use the Cauchy--Schwarz inequality for the inner product $\Tr[\rho X^\dagger Y]$:
\begin{equation}
    \big|\Tr[\rho X^\dagger Y]\big|^2\leq\Tr[\rho X^\dagger X]\,\Tr[\rho Y^\dagger Y].
    \label{eq:app_cs}
\end{equation}

\subsection{Response of the MFG state}
\label{app:mfg_response}

We need $\partial_\delta\tilde\omega$ for the family $H_{\rm tot}+\delta\kappa$. The route taken here differentiates the global Gibbs state and traces out the bath only at the end, so that the mean-force Hamiltonian is never constructed.  For operators $X$ and $Y$, the derivative of the exponential is~\cite{metrology_thermalequilibrium,Scandi2025QFIDynamicalNature}
\begin{equation}
    \partial_\delta e^{X+\delta Y}\big|_{\delta=0}=\int_0^1ds\;e^{sX}\,Y\,e^{(1-s)X},
    \label{eq:app_expder}
\end{equation}
which reduces to $Ye^X$ only if $[X,Y]=0$. We take $X=-\beta H_{\rm tot}$ and $Y=-\beta\kappa$. The trace of Eq.~\eqref{eq:app_expder} gives $\partial_\delta\ln Z=-\beta\langle\kappa\rangle$. Since $\int_0^1ds\,\omega_{\rm tot}^s\,\omega_{\rm tot}^{1-s}=\omega_{\rm tot}$, the normalization centers the perturbation:
\begin{equation}
    \partial_\delta\,\omega(\beta,H_{\rm tot}+\delta\kappa)\big|_{\delta=0}
    =-\beta\int_0^1ds\;\omega_{\rm tot}^s\,\Delta\kappa\,\omega_{\rm tot}^{1-s}.
    \label{eq:app_gibbs_derivative}
\end{equation}
The partial trace of Eq.~\eqref{eq:app_gibbs_derivative} is $\partial_\delta\tilde\omega=-\beta R_\kappa$, with $R_\kappa$ defined in Eq.~\eqref{eq:response_operator}. Hence, for any system observable $A$,
\begin{equation}
    \partial_\delta\Tr[\tilde\omega A]=-\beta\Tr[AR_\kappa]=-\xi_{A\kappa}.
    \label{eq:app_response}
\end{equation}
To obtain the second form of Eq.~\eqref{eq:static_susceptibility}, replace $A$ by $\Delta A$, which is allowed because $\Tr R_\kappa=\Tr[\tilde\omega\Delta\kappa]=0$. Then cycle $\omega_{\rm tot}^{1-s}$ to the front of the trace and write $\omega_{\rm tot}^s=e^{-s\beta H_{\rm tot}}/Z^s$:
\begin{equation}
    \xi_{A\kappa}
    =\beta\int_0^1ds\;\Tr\big[\omega_{\rm tot}^{1-s}\,\Delta A\,\omega_{\rm tot}^{s}\,\Delta\kappa\big]
    =\int_0^\beta d\lambda\;\Tr\big[\omega_{\rm tot}\,e^{\lambda H_{\rm tot}}\Delta A\,e^{-\lambda H_{\rm tot}}\,\Delta\kappa\big],
    \label{eq:app_kubo_mori}
\end{equation}
with $\lambda=\beta s$. The operator $e^{\lambda H_{\rm tot}}\Delta A\,e^{-\lambda H_{\rm tot}}$ is $\Delta A(-i\hbar\lambda)$. The first form is $\beta$ times the canonical (Kubo--Mori) correlation of $A$ and $\kappa$~\cite{KuboTodaHashitsume1991}. It is invariant under $s\to1-s$, so $\xi_{A\kappa}=\xi_{\kappa A}$. Equation~\eqref{eq:app_kubo_mori} is what the rest of the appendix works with: the response of the MFG state to a perturbation of $H_S$ is a two-point equilibrium correlation of the global state, with no transient dynamics and no mean-force Hamiltonian anywhere in it.

\subsection{Equilibrium correlations and spectral representation}
\label{app:spectral}

Equation~\eqref{eq:app_kubo_mori} expresses the susceptibility through an imaginary-time correlation. Here we rewrite it in terms of real-time equilibrium correlations and their spectrum, which are the standard objects of linear-response theory. The outcome is that the susceptibility $\xi_{\kappa\kappa}$ and the variance $\operatorname{Var}_{\tilde\omega}(\kappa)$, which together fix the bounds of Sec.~\ref{subsec:mfg_qfi}, are two weighted integrals of the same fluctuation spectrum. They differ only by the Planckian filter, which is the origin of the Planckian window in Eq.~\eqref{eq:susceptibility_correlations}.

We start from the symmetrized equilibrium correlation function of two system observables,
\begin{equation}
    C_{A\kappa}(t)=\frac12\Tr\big[\omega_{\rm tot}\{\Delta A(t),\Delta\kappa\}\big],
    \qquad
    X(t)=e^{iH_{\rm tot}t/\hbar}Xe^{-iH_{\rm tot}t/\hbar}.
    \label{eq:app_correlation}
\end{equation}
Although $A$ and $\kappa$ act on the system, they evolve with the full Hamiltonian, so $C_{A\kappa}$ contains the effect of the bath on the system fluctuations.
Its equal-time value, $C_{A\kappa}(0)=\frac12\Tr[\tilde\omega\{\Delta A,\Delta\kappa\}]$, involves only the MFG state; in particular, $C_{\kappa\kappa}(0)=\operatorname{Var}_{\tilde\omega}(\kappa)$.

The frequency content of these fluctuations is described by the spectrum. With angular frequencies and the two-sided convention
\begin{equation}
    S_{A\kappa}(\nu)=\frac1{2\pi}\int dt\,e^{i\nu t}\,C_{A\kappa}(t),
    \qquad
    C_{A\kappa}(t)=\int d\nu\,e^{-i\nu t}\,S_{A\kappa}(\nu),
    \label{eq:app_fourier}
\end{equation}
the total spectral weight is the equal-time correlation, $\int d\nu\,S_{A\kappa}(\nu)=C_{A\kappa}(0)$. To identify how each frequency contributes, we insert the eigenstates of the total Hamiltonian, $H_{\rm tot}|m\rangle=E_m|m\rangle$, with populations $q_m=\langle m|\omega_{\rm tot}|m\rangle$. Each pair of levels contributes a line at its Bohr frequency, weighted by the arithmetic mean of the two populations:
\begin{equation}
    S_{A\kappa}(\nu)=\sum_{mn}\frac{q_m+q_n}{2}\,(\Delta A)_{mn}(\Delta\kappa)_{nm}\;\delta\!\Big(\nu-\frac{E_n-E_m}{\hbar}\Big).
    \label{eq:app_spectrum}
\end{equation}
For $A=\kappa$, the weights $|(\Delta\kappa)_{mn}|^2$ are nonnegative. Hence $S_{\kappa\kappa}$ is a nonnegative, even measure whose total weight is $\operatorname{Var}_{\tilde\omega}(\kappa)$. It is the fluctuation spectrum of a system observable in the interacting equilibrium state, and it should not be confused with the spectral density of the bath.

The susceptibility involves the same matrix elements, but with a different weight. Inserting the same eigenstates in the first form of Eq.~\eqref{eq:app_kubo_mori} gives
\begin{equation}
    \xi_{A\kappa}=\beta\sum_{mn}\mu_{mn}\,(\Delta A)_{mn}(\Delta\kappa)_{nm},
    \qquad
    \mu_{mn}=\int_0^1ds\,q_m^{1-s}q_n^{s}=\frac{q_m-q_n}{\ln q_m-\ln q_n},
    \label{eq:app_km_weights}
\end{equation}
where the arithmetic mean of the populations is replaced by their logarithmic mean. The two weights are related by a function of the Bohr frequency alone. With $x=\beta(E_n-E_m)$, so that $q_n/q_m=e^{-x}$,
\begin{equation}
    \frac{2\mu_{mn}}{q_m+q_n}=\frac{2(1-e^{-x})}{x(1+e^{-x})}=f(x),
    \qquad
    f(x)=\frac{\tanh(x/2)}{x/2},
    \label{eq:app_ratio}
\end{equation}
with $f(0)=1$. Since each line of Eq.~\eqref{eq:app_spectrum} sits at $x=\beta\hbar\nu$, the susceptibility is a filtered integral of the spectrum,
\begin{equation}
    \xi_{A\kappa}=\beta\int d\nu\;S_{A\kappa}(\nu)\,f(\beta\hbar\nu),
    \label{eq:app_xi_S}
\end{equation}
which is the second line of Eq.~\eqref{eq:susceptibility_correlations}. The function $f(\beta \hbar \nu)=\tanh(\beta \hbar \nu/2)/(\beta \hbar \nu/2)$ is the Planckian filter introduced in Sec.~\ref{subsec:mfg_response}. It equals one for slow fluctuations, $\beta\hbar|\nu|\ll1$, and suppresses fast ones as $2/(\beta\hbar|\nu|)$. For $A=\kappa$, the variance is thus the unfiltered weight of $S_{\kappa\kappa}$, and the susceptibility its filtered weight.

Finally, the filter can be moved to the time domain. Since $f$ is even, inserting $S_{A\kappa}$ from Eq.~\eqref{eq:app_fourier} into Eq.~\eqref{eq:app_xi_S} gives the first line of Eq.~\eqref{eq:susceptibility_correlations}, $\xi_{A\kappa}=\beta\int dt\,w_\beta(t)\,C_{A\kappa}(t)$, with the Planckian window
\begin{equation}
    w_\beta(t)=\frac1{2\pi}\int d\nu\,e^{-i\nu t}f(\beta\hbar\nu)
    =\frac{2}{\pi\beta\hbar}\ln\coth\frac{\pi|t|}{2\beta\hbar}.
    \label{eq:app_kernel}
\end{equation}
Note that this   window is positive, even and normalized, $\int dt\,w_\beta(t)=f(0)=1$. It has an integrable logarithmic singularity at $t=0$, and its slowest component decays as $e^{-\pi|t|/(\beta\hbar)}$. The susceptibility is therefore the equal-time correlation smoothed over the Planckian window: correlations that persist over it contribute fully, whereas those that decay faster are averaged out. The same window represents the SLD of a Gibbs state as a time average of the perturbing operator~\cite{MengShi2026ThermodynamicUncertainty}.

The real-time form also gives the bracket of Sec.~\ref{subsec:mfg_qfi} directly. For $A=\kappa$, $C_{\kappa\kappa}(t)={\rm Re}\Tr[\omega_{\rm tot}\Delta\kappa(t)\Delta\kappa]$. Equation~\eqref{eq:app_cs} with $\rho=\omega_{\rm tot}$,  $X=\Delta\kappa(t)$ and $Y=\Delta\kappa$, together with stationarity, $\Tr[\omega_{\rm tot}\Delta\kappa(t)^2]=\Tr[\omega_{\rm tot}\Delta\kappa^2]$, gives
\begin{equation}
    |C_{\kappa\kappa}(t)|\leq C_{\kappa\kappa}(0)=\operatorname{Var}_{\tilde\omega}(\kappa).
    \label{eq:app_cs_correlation}
\end{equation}
Since $w_\beta$ is positive and normalized, $\xi_{\kappa\kappa}=\beta\int dt\,w_\beta(t)C_{\kappa\kappa}(t)\leq\beta\operatorname{Var}_{\tilde\omega}(\kappa)$. The susceptibility is also strictly positive whenever $\operatorname{Var}_{\tilde\omega}(\kappa)>0$, because $S_{\kappa\kappa}\geq0$ carries the total weight $\operatorname{Var}_{\tilde\omega}(\kappa)$ and $f>0$ in Eq.~\eqref{eq:app_xi_S}. Hence $r_\kappa\in(0,1]$ in Eq.~\eqref{eq:r_kappa}.

The upper value $r_\kappa=1$ is reached only for conserved perturbations. Because $w_\beta(t)>0$ at every $t$, equality in $\xi_{\kappa\kappa}\leq\beta\operatorname{Var}_{\tilde\omega}(\kappa)$ requires $C_{\kappa\kappa}(t)=C_{\kappa\kappa}(0)$ for all $t$. Since $\Tr[\omega_{\rm tot}(\Delta\kappa(t)-\Delta\kappa)^2]=2[C_{\kappa\kappa}(0)-C_{\kappa\kappa}(t)]$ and $\omega_{\rm tot}$ has full rank, this holds if and only if $\Delta\kappa(t)=\Delta\kappa$ for all $t$, i.e., $[H_{\rm tot},\kappa]=0$. In spectral terms, all the weight of $S_{\kappa\kappa}$ then sits at $\nu=0$, where $f=1$.

\subsection{Proof of Proposition~\ref{prop:mfg_qfi}}
\label{app:proof_prop}

Proposition~\ref{prop:mfg_qfi} identifies $\mathcal F^{\rm th}_\kappa$ with the largest value of $\xi_{A\kappa}^2/\operatorname{Var}_{\tilde\omega}(A)$ over system observables. The proof applies Eq.~\eqref{eq:app_cs} at $\rho=\tilde\omega$, once the susceptibility is recognized as an overlap between $\Delta A$ and the SLD.

Let $L_\kappa$ be the SLD of the MFG family, $\partial_\delta\tilde\omega=\frac12\{\tilde\omega,L_\kappa\}$, so that $\Tr[\tilde\omega L_\kappa]=0$ and $\mathcal F^{\rm th}_\kappa=\Tr[\tilde\omega L_\kappa^2]$~\cite{Paris2009QuantumEstimation}. For a system observable $A$, Eq.~\eqref{eq:app_response} and $\Tr[\partial_\delta\tilde\omega]=0$ give
\begin{equation}
    -\xi_{A\kappa}=\Tr[\Delta A\,\partial_\delta\tilde\omega]={\rm Re}\,\Tr[\tilde\omega\,\Delta A\,L_\kappa].
\end{equation}
Equation~\eqref{eq:app_cs} with $\rho=\tilde\omega$, $X=\Delta A$, and $Y=L_\kappa$ yields $\xi_{A\kappa}^2\leq\operatorname{Var}_{\tilde\omega}(A)\,\mathcal F^{\rm th}_\kappa$, with equality for $A=L_\kappa$~\cite{BraunsteinCaves1994,Paris2009QuantumEstimation}. This proves Eq.~\eqref{eq:qfi_variational}, and the choice $A=\kappa$ gives the first inequality of Eq.~\eqref{eq:qfi_hierarchy}.

We restate the chain of Eq.~\eqref{eq:qfi_hierarchy} here,
\begin{equation}
    \frac{\xi_{\kappa\kappa}^2}{\operatorname{Var}_{\tilde\omega}(\kappa)}
    \leq\mathcal F^{\rm th}_\kappa
    \leq\beta\,\xi_{\kappa\kappa}
    \leq\beta^2\operatorname{Var}_{\tilde\omega}(\kappa)
    \leq\frac{\beta^2\|\kappa\|^2}{4},
    \tag{\ref{eq:qfi_hierarchy}}
\end{equation}
and prove its four inequalities in turn. The first is the Cauchy--Schwarz inequality just used, Eq.~\eqref{eq:qfi_variational} evaluated at $A=\kappa$. The second compares two monotone Riemannian metrics of the global Gibbs family $\omega(\beta,H_{\rm tot}+\delta\kappa)$: its Kubo--Mori metric is $\beta\xi_{\kappa\kappa}$, by the first form of Eq.~\eqref{eq:app_kubo_mori}, and by Petz's classification of monotone metrics the SLD metric is the smallest of them~\cite{Petz1996MonotoneMetrics, MillerAnders2018EnergyTemperature, Tonchev2021MonotoneMetrics}, so that $\mathcal F[\omega_{\rm tot}]\leq\beta\xi_{\kappa\kappa}$. The MFG family inherits this bound because the SLD metric does not increase under the partial trace, $\mathcal F^{\rm th}_\kappa\leq\mathcal F[\omega_{\rm tot}]$, while $\xi_{\kappa\kappa}$ and $\operatorname{Var}_{\tilde\omega}(\kappa)$ take the same value in $\omega_{\rm tot}$ and in $\tilde\omega$, since $\kappa$ acts only on the system. The third was proved in App.~\ref{app:spectral}, below Eq.~\eqref{eq:app_cs_correlation}: the susceptibility is the equal-time correlation averaged over the positive, normalized Planckian window, and $|C_{\kappa\kappa}(t)|\leq C_{\kappa\kappa}(0)$. The fourth is the usual variance bound for a bounded observable~\cite{Boixo2007GeneralizedLimits,metrology_thermalequilibrium}: shifting $\kappa$ by the midpoint $c$ of its spectrum, $\operatorname{Var}_{\tilde\omega}(\kappa)\leq\langle(\kappa-c)^2\rangle\leq\|\kappa\|^2/4$.

\subsection{Qubit targets}
\label{app:qubit}

This subsection derives case~(iii) of Sec.~\ref{subsec:optimal_perturbations}. The idea is that, for a qubit, the QFI is a quadratic form in the coefficients of $\kappa$ while the constraint $\|\kappa\|=1$ of Eq.~\eqref{eq:local_hamiltonian_perturbation} is the unit sphere, so that the optimization over perturbations becomes a $3\times3$ eigenvalue problem.

\paragraph*{Basis expansion.} Expand the perturbation and the observable in a basis of traceless system observables $\{A_i\}$, $\kappa=\sum_jk_jA_j+c\,\mathbb{I}$ and $A=\sum_ia_iA_i$; constants do not contribute to $\xi_{A\kappa}$ or to $\operatorname{Var}_{\tilde\omega}(A)$. Then $\xi_{A\kappa}=\beta\,\bm a^\top M \bm k$ and $\operatorname{Var}_{\tilde\omega}(A)=\bm a^\top \Sigma \bm a$, with $M$ and $\Sigma$ as in Sec.~\ref{subsec:optimal_perturbations}. The Cauchy--Schwarz inequality for the inner product $\bm  a^\top \Sigma \bm a'$ gives
\begin{equation}
    \max_{\bm a}\frac{\big(\beta\,\bm a^\top M\bm k\big)^2}{\bm a^\top \Sigma \bm a}=\beta^2\,\bm k^\top M\Sigma^{-1}M\bm k,
    \qquad
    \bm a^*\propto \Sigma^{-1}M\bm k.
    \label{eq:app_basis}
\end{equation}
If $\{A_i\}$ spans all traceless observables, the SLD is of this form, and Eq.~\eqref{eq:app_basis} equals $\mathcal F^{\rm th}_\kappa$ by Eq.~\eqref{eq:qfi_variational}.

\paragraph*{Qubit.} For a qubit, the Pauli matrices form such a basis. With $\tilde\omega=\frac12(\mathbb{I}+\bm b\cdot\bm\sigma)$ and $\frac12\{\sigma_i,\sigma_j\}=\delta_{ij}\mathbb{I}$, one finds $\Sigma=\mathbb{I}-\bm b\bm b^\top$. For $\kappa=\frac12\bm n\cdot\bm\sigma$, Eq.~\eqref{eq:app_response} gives $\partial_\delta b_i=-\xi_{\sigma_i\kappa}=\sum_jJ_{ij}n_j$, with $J_{ij}=-\xi_{\sigma_i\sigma_j}/2$, i.e., $M=-2J/\beta$ and $k=\bm n/2$. Equation~\eqref{eq:app_basis} becomes
\begin{equation}
    \mathcal F^{\rm th}_\kappa=\bm n^\top K\bm n,
    \qquad
    K=J\Sigma^{-1}J,
    \qquad
    \Sigma^{-1}=\mathbb{I}+\frac{\bm b\bm b^\top}{1-|\bm b|^2}.
    \label{eq:app_qubit_qfi}
\end{equation}
Writing $\partial_\delta\bm b=J\bm n$, this is the Bloch-vector form of the QFI quoted in case~(iii)~\cite{Zhong2013BlochFisher}. The norm is $\|\kappa\|=|\bm n|$, and every nontrivial projector is a shifted perturbation of this form, $\frac12(\mathbb{I}+\bm n\cdot\bm\sigma)$ with $|\bm n|=1$. Maximizing the Rayleigh quotient $\bm n^\top K\bm n$ over unit vectors gives $\lambda_{\max}(K)$, i.e., Eq.~\eqref{eq:qubit_chi}, with the optimal perturbation along the top eigenvector $\bm n^*$ of $K$ and the optimal observable $A^*=(\Sigma^{-1}J\bm n^*)\cdot\bm\sigma$.

\paragraph*{Longitudinal and transverse channels.} If $J\bm b=J_\parallel\bm b$, the symmetric matrix $J$ has an orthonormal eigenbasis $\{\hat{\bm b},\bm e_1,\bm e_2\}$, with $J\bm e_k=J_{\perp,k}\bm e_k$. Since $\Sigma^{-1}\hat{\bm b}=\hat{\bm b}/(1-|\bm b|^2)$ and $\Sigma^{-1}\bm e_k=\bm e_k$, the matrix $K$ is diagonal in the same basis, with eigenvalues $J_\parallel^2/(1-|\bm b|^2)$ and $J_{\perp,k}^2$, which gives Eq.~\eqref{eq:qubit_channels}. For each of these directions $\bm u$, the optimal observable $\Sigma^{-1}J\bm u\cdot\bm\sigma$ is proportional to $\bm u\cdot\bm\sigma$. The SLD is therefore proportional to $\kappa$ up to a multiple of $\mathbb{I}$, and Eq.~\eqref{eq:self_response_bound} with $A=\kappa$ is exact.

A sufficient condition is a symmetry $U=R\otimes U_B$ of $H_{\rm tot}$, with $R$ a $\pi$ rotation of the qubit about an axis $\hat{\bm m}$. Since $U$ commutes with $\omega_{\rm tot}$, $\tilde\omega=R\tilde\omega R^\dagger$, so that $\bm b$ has no component perpendicular to $\hat{\bm m}$. Moreover, the first form of Eq.~\eqref{eq:app_kubo_mori} gives $\xi_{U^\dagger XU,\,U^\dagger YU}=\xi_{XY}$. Since $R$ leaves $\sigma_{\hat m}=\hat{\bm m}\cdot\bm\sigma$ invariant and reverses the perpendicular Pauli operators $\sigma_\perp$, this implies $\xi_{\sigma_{\hat m}\sigma_\perp}=-\xi_{\sigma_{\hat m}\sigma_\perp}=0$: $\hat{\bm m}$ is an eigenvector of $J$. If $H_{\rm tot}$ is invariant under all rotations about $\hat{\bm m}$, the same argument shows that $J$ commutes with them, so that $J_{\perp,1}=J_{\perp,2}$.

\paragraph*{Weak coupling.} For $\tilde\omega=\omega(\beta,\bar H_S)$ with $\bar H_S=\frac\Delta2\sigma_z$, $\bm b=-\tanh(x)\hat{\bm z}$, with $x=\beta\Delta/2$. The perturbation $\frac12\bm n\cdot\bm\sigma$ shifts the field of $\bar H_S$ from $\Delta\hat{\bm z}$ to $\Delta\hat{\bm z}+\delta\bm n$. Along $\hat{\bm z}$, it changes the length of $\bm b$, $J_\parallel=-\frac\beta2{\rm sech}^2x$; perpendicular to $\hat{\bm z}$, it rotates $\bm b$ by the angle $\delta/\Delta$, $J_\perp=-\tanh(x)/\Delta$. With $1-|\bm b|^2={\rm sech}^2x$, Eq.~\eqref{eq:qubit_channels} gives the longitudinal value $\frac12{\rm sech}\,x$ and the transverse value $\tanh(x)/(2x)$, and the latter is larger because $\sinh x\geq x$. In Sec.~\ref{sec:spin_boson}, the symmetric point $\theta=\pi/2$ realizes the eigenvector case at every coupling, and the pinched state of Eq.~\eqref{eq:sb_usc} retains only the longitudinal channel.

\section{Fermionic models: exact solution, equilibrium sensitivity, and speed limits}
\label{app:fermionic}

This appendix collects the exact solution of the model in Eq.~\eqref{eq:fermionic_hamiltonian}, the QFI of its MFG states, and the comparison with speed limits. We set $\hbar=1$  in the first two subsections, restoring it from App.~\ref{app:dark_family} on, and write $H_S=\sum_{mn}h_{mn}d_m^\dagger d_n$ with a Hermitian one-particle matrix $h$, which covers both $h={\rm diag}(\epsilon_1,\dots,\epsilon_N)$ and the quadratic perturbations of Sec.~\ref{subsec:dqd}.

\subsection{Exact dynamics and MFG states} The derivation follows the single-level case of Ref.~\cite{Rolandi2023finitetimelandauer}. The new element for $N$ levels is that all of them couple to the same reservoir channel. Eliminating the reservoir from the Heisenberg equations and taking the wide-band limit for $t\gg1/\Lambda$ gives the quantum Langevin equation
\begin{equation}
    \dot{\mathbf d}(t)=\Gamma\,\mathbf d(t)-i\eta(t)\,\bm u,
    \qquad
    \eta(t)=g\sum_kt_k\,e^{-i\Omega_kt}\,b_k(0),
    \label{eq:app_langevin}
\end{equation}
with $\mathbf d=(d_1,\dots,d_N)^{\rm T}$ and $\Gamma=-ih-(g^2/2)\,\bm u\bm u^{\rm T}$, as in Eq.~\eqref{eq:damping_matrix}. Because every level feels the same noise $\eta(t)$, the damping in $\Gamma$ is collective, $\propto \bm u \bm u^{\rm T}$, rather than diagonal. For a reservoir initially in $\omega(\beta,H_B)$, $\langle\eta^\dagger(s)\eta(s')\rangle=g^2\int\frac{d\Omega}{2\pi}f_\beta(\Omega)\,e^{i\Omega(s-s')}$ with $f_\beta(\Omega)=(1+e^{\beta\Omega})^{-1}$. Solving Eq.~\eqref{eq:app_langevin} for a factorized initial state gives
\begin{equation}
    C(t)=D(t)\,C(0)\,D(t)^{\dagger}
    +g^2\!\int\!\frac{d\Omega}{2\pi}\,f_\beta(\Omega)\,
    \bm r_\Omega(t)\,\bm r_\Omega(t)^{\dagger},
    \qquad
    \bm r_\Omega(t)=\int_0^t\!ds\,D(t-s)\,\bm u\,e^{-i\Omega s},
    \label{eq:app_correlation_dynamics}
\end{equation}
with $D(t)=e^{\Gamma t}$. The first term carries the memory of the initial state and decays with the eigenvalues of $\Gamma$. The second term, weighted by the Fermi function, is the contribution of the reservoir and fixes the long-time state.

\paragraph*{Stationary state and MFG state.} If all eigenvalues of $\Gamma$ have negative real parts, $D(t)\to0$ and $\bm r_\Omega(t)\to e^{-i\Omega t}\bm r_\Omega$ with $\bm r_\Omega=-(\Gamma+i\Omega)^{-1}\bm u=iG(\Omega)\bm u$. Here $G(\Omega)=(\Omega-i\Gamma)^{-1}$, and $i\Gamma=h-i(g^2/2)\,\bm u \bm u^{\rm T}$ is the non-Hermitian effective Hamiltonian of the levels. The phase cancels in Eq.~\eqref{eq:app_correlation_dynamics}, so that
\begin{equation}
    C^{\rm ss}=g^2\int\frac{d\Omega}{2\pi}\,f_\beta(\Omega)\,G(\Omega)\, \bm u \bm u^{\rm T}\,G(\Omega)^\dagger.
    \label{eq:app_stationary}
\end{equation}
To compare with the MFG state, we diagonalize the quadratic total Hamiltonian, $H_{\rm tot}=\sum_kE_k\,c_k^\dagger c_k$, and expand the level operators in its single-particle eigenstates $|k\rangle=c_k^\dagger|0\rangle$, $d_n=\sum_k\langle0|d_n|k\rangle\,c_k$. In the global Gibbs state $\langle c_k^\dagger c_{k'}\rangle=\delta_{kk'}f_\beta(E_k)$, so
\begin{equation}
    C^{\rm th}_{mn}=\Tr\!\big[\omega(\beta,H_{\rm tot})\,d_n^\dagger d_m\big]=\int\frac{d\Omega}{2\pi}\,f_\beta(\Omega)\,\varphi_{mn}(\Omega),
    \qquad
    \varphi_{mn}(\Omega)=2\pi\sum_k\delta(\Omega-E_k)\,\langle0|d_m|k\rangle\langle k|d_n^\dagger|0\rangle.
    \label{eq:app_gibbs_correlation}
\end{equation}
The spectral function $\varphi(\Omega)$ gives the weight on the levels of the single-particle eigenstates with energy $\Omega$. The same expansion gives the amplitude for a particle created on level $n$ to be found on level $m$ after a time $t$, $\langle0|d_m(t)d_n^\dagger|0\rangle=\int\frac{d\Omega}{2\pi}\varphi_{mn}(\Omega)\,e^{-i\Omega t}$. Since the noise in Eq.~\eqref{eq:app_langevin} annihilates the vacuum, this amplitude also follows from the Langevin solution as $D_{mn}(t)$. Equating the one-sided Fourier transforms of the two expressions, $\int_0^\infty dt\,e^{i\Omega t}D(t)=iG(\Omega)$ and $\frac12\varphi(\Omega)+\frac{i}{2\pi}\mathcal P\!\int d\Omega'\,\varphi(\Omega')/(\Omega-\Omega')$, gives
\begin{equation}
    \varphi(\Omega)=i\big[G(\Omega)-G(\Omega)^\dagger\big]=g^2\,G(\Omega)\,\bm u \bm u^{\rm T}\,G(\Omega)^\dagger,
    \label{eq:app_spectral_function}
\end{equation}
where the second form follows from $i\Gamma-(i\Gamma)^\dagger=-ig^2\bm u \bm u^{\rm T}$. Comparing Eqs.~\eqref{eq:app_stationary} and \eqref{eq:app_gibbs_correlation} then gives $C^{\rm ss}=C^{\rm th}$. Since both states are number-conserving Gaussian states, $\lim_{t\to\infty}\rho_S(t)=\tilde\omega(\beta,H_S)$. For the DQD at $\Delta=0$, one eigenvalue of $\Gamma$ is purely imaginary, the dark mode keeps its initial occupation, and the stationary state depends on the preparation.

\paragraph*{Closed form of the MFG correlation matrix.} It remains to evaluate $\int\frac{d\Omega}{2\pi}f_\beta(\Omega)\varphi(\Omega)$ with $\varphi(\Omega)=i[(\Omega-i\Gamma)^{-1}-(\Omega-(i\Gamma)^\dagger)^{-1}]$. For a single level, $i\Gamma=z=\epsilon-ig^2/2$ is a number with ${\rm Im}\,z<0$, and
\begin{equation}
    \int\frac{d\Omega}{2\pi}\,f_\beta(\Omega)\,i\left[\frac{1}{\Omega-z}-\frac{1}{\Omega-z^*}\right]
    =\frac12-\frac1\pi\,{\rm Im}\,\psi\!\left(\frac12+\frac{i\beta z}{2\pi}\right),
    \label{eq:app_scalar_digamma}
\end{equation}
which is the occupation in Eq.~\eqref{eq:rlm_chi}~\cite{Abiuso2025PlanckianBound}. Equation~\eqref{eq:app_scalar_digamma} follows from $\Phi(z)\equiv\int d\Omega\,f_\beta(\Omega)(\Omega-z)^{-1}=\psi(\frac12+\frac{i\beta z}{2\pi})+c$, since its left-hand side equals $(i/2\pi)[\Phi(z)-\Phi(z)^*]$. The constant $c$ is independent of $z$: its real part diverges logarithmically with the high-energy cutoff and cancels in the difference, while its imaginary part, $-\pi/2$, produces the term $1/2$. For $N$ levels, $i\Gamma$ is diagonalizable, $i\Gamma=S\,Z\,S^{-1}$ with $Z={\rm diag}(z_1,\dots,z_N)$, with all eigenvalues in the lower half-plane, except at isolated exceptional points such as $g^2=|\Delta|$ in the DQD, which follow by continuity. Since $i\Gamma$ is not normal, $S$ is not unitary. Nevertheless, $(\Omega-i\Gamma)^{-1}=S(\Omega-Z)^{-1}S^{-1}$, so the frequency integral acts on each eigenvalue separately and gives $\int d\Omega\,f_\beta(\Omega)(\Omega-i\Gamma)^{-1}=S\,\Phi(Z)\,S^{-1}=\psi(\frac12\mathbb{I}-\frac{\beta}{2\pi}\Gamma)+c\,\mathbb{I}$. Therefore,
\begin{equation}
    \int\frac{d\Omega}{2\pi}\,f_\beta(\Omega)\,\varphi(\Omega)=\frac12\,\mathbb{I}-\frac{1}{2\pi i}\left[\psi\!\left(\frac12\,\mathbb{I}-\frac{\beta}{2\pi}\Gamma\right)-{\rm h.c.}\right].
    \label{eq:app_closed_form}
\end{equation}
By Eq.~\eqref{eq:app_gibbs_correlation}, $C^{\rm th}$ is Eq.~\eqref{eq:app_closed_form} itself, which is Eq.~\eqref{eq:mfg_correlation} with $\hbar$ restored.

\subsection{Quantum Fisher information of the MFG states} A number-conserving Gaussian state is fully determined by its $N\times N$ correlation matrix, and so is its QFI. For a quadratic perturbation $\kappa=\sum_{mn}k_{mn}d_m^\dagger d_n$, i.e., $h\to h+\delta k$, the total Hamiltonian stays quadratic and the perturbed targets stay Gaussian. Writing $C^{\rm th}=U\,{\rm diag}(\gamma_1,\dots,\gamma_N)\,U^\dagger$, where the columns of $U$ define the natural orbitals of the MFG state and $\gamma_i$ their occupations, the QFI is~\cite{carolloSymmetricLogarithmicDerivative2018}
\begin{equation}
    \mathcal F^{\rm th}_\kappa=\sum_i\frac{(\partial_\delta\gamma_i)^2}{\gamma_i(1-\gamma_i)}
    +2\sum_{i\neq j}\frac{(\gamma_i-\gamma_j)^2}{\gamma_i+\gamma_j-2\gamma_i\gamma_j}\,\big|(U^\dagger\partial_\delta U)_{ij}\big|^2.
    \label{eq:app_gaussian_qfi}
\end{equation}
Equation~\eqref{eq:app_gaussian_qfi} has three useful properties.

(i)~It requires only $C^{\rm th}$ and $\partial_\delta C^{\rm th}$, which follow from Eq.~\eqref{eq:mfg_correlation}. Neither the $2^N$-dimensional density matrix nor the reservoir enters explicitly.

(ii)~It separates two sources of equilibrium sensitivity. The first term is the classical Fisher information of the natural-orbital occupations, i.e., the information accessible by measuring the occupations of the natural orbitals. The second term arises from the rotation of the natural orbitals, i.e., from coherences between them, and it vanishes when $\partial_\delta C^{\rm th}$ commutes with $C^{\rm th}$. For a single level only the first term is present, and Eq.~\eqref{eq:app_gaussian_qfi} reduces to Eq.~\eqref{eq:rlm_chi}. In the DQD at strong coupling, the natural orbitals approach the bright and dark modes, which are fixed by the coupling vector $\bm u$ and do not depend on $\bar\epsilon$, with occupations $1/2$ and $f_\beta(\bar\epsilon)$. The QFI then reduces to the classical information carried by the dark occupation. Since both $N_S$ and $n_-$ shift the dark-mode energy by $\delta$, both give the same limit, $\mathcal F^{\rm th}_{N_S},\mathcal F^{\rm th}_{n_-}\to\beta^2f_\beta(\bar\epsilon)\big[1-f_\beta(\bar\epsilon)\big]$. With the normalization $\|\kappa\|=1$ of Eq.~\eqref{eq:local_hamiltonian_perturbation}, the common shift enters as $N_S/2$, for which homogeneity gives $\mathcal F^{\rm th}_{N_S/2}=\mathcal F^{\rm th}_{N_S}/4$; this is the factor of two in $\widetilde\chi$ discussed in Sec.~\ref{subsec:dqd}.

(iii)~Because the MFG states are Gaussian, their SLD is a quadratic, number-conserving operator, $L_\kappa=\sum_{mn}\ell_{mn}(d_m^\dagger d_n-C^{\rm th}_{\,{nm}})$, whose coefficients follow from $C^{\rm th}$ and $\partial_\delta C^{\rm th}$~\cite{carolloSymmetricLogarithmicDerivative2018}. Since the SLD attains the maximum in Eq.~\eqref{eq:qfi_variational}, measuring the mean of a single one-body observable, a combination of occupations and interlevel coherences, recovers the full QFI. In the strongly coupled DQD, the natural orbitals no longer depend on $\bar\epsilon$ and only the dark occupation responds, so $L_\kappa\propto n_--\langle n_-\rangle$ for both $\kappa=N_S$ and $\kappa=n_-$. This is the readout identified in Fig.~\ref{fig:local_sensitivity}(c).

\subsection{Dark-mode family}\label{app:dark_family}
\begin{figure}[t]
    \centering
    \includegraphics[width=\textwidth]{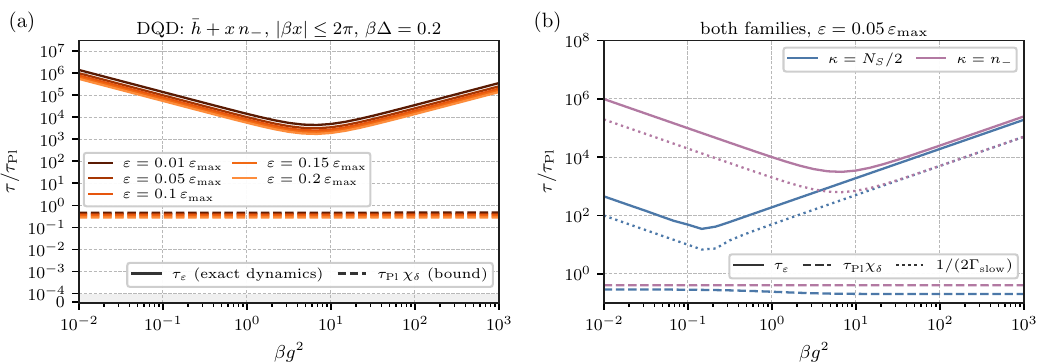}
    \caption{\textbf{Dark-mode family.}
    (a)~Thermalization time $\tau_\varepsilon$ (solid) and bound $\tau_{\rm Pl}\chi_\delta$ (dashed) for the family $h={\rm diag}(1.2,1.0)\,k_BT+x\,|{-}\rangle\langle{-}|$, i.e., $\kappa=n_-$, with $|\beta x|\leq2\pi$, $\beta\Delta=0.2$, and initial occupations $(n_1,n_2)=(1,0)$, for the accuracies of Fig.~\ref{fig:finite_error}. The vertical axis is linear below $10^{-4}$.
    (b)~Common-shift and dark-mode families at $\varepsilon=0.05\,\varepsilon_{\max}$: $\tau_\varepsilon$ (solid), $\tau_{\rm Pl}\chi_\delta$ (dashed), and the occupation relaxation time $1/(2\Gamma_{\rm slow})$ of the slowest member of the family, obtained from the eigenvalues of $\Gamma$ (dotted). Wide-band dynamics.}
    \label{fig:dark_comparison}
\end{figure}

In the dot basis, the common shift and the dark-mode occupation correspond to $k=\mathbb{I}/2$ and $k=(\mathbb{I}-\sigma_x)/2$, respectively, and $\bm u \bm u^{\rm T}=\mathbb{I}+\sigma_x$. Along the common shift, Eq.~\eqref{eq:damping_matrix} gives $\Gamma(x)=\Gamma(0)-ix\,\mathbb{I}/(2\hbar)$: all eigenvalues are shifted by the same imaginary amount, so that every member has the decay rates of $\bar H_S$. Along the dark-mode direction, $h$ acquires the additional tunneling term $-(x/2)\sigma_x$, which combines with the coupling term $-(g^2/2\hbar)\sigma_x$ of $\Gamma$. The eigenvalues become
\begin{equation}
    \lambda_\pm(x)=-\frac{i}{\hbar}\Big(\bar\epsilon+\frac{x}{2}\Big)-\frac{g^2}{2\hbar}\pm\frac{1}{2\hbar}\sqrt{(g^2-ix)^2-\Delta^2},
    \label{eq:app_dark_eigs}
\end{equation}
which reduce to those of Sec.~\ref{subsec:dqd} at $x=0$. The decay rates ${\rm Re}\,\lambda_\pm$ now depend on $x$: the slow rate falls off with $|x|$ on the scale $\max(g^2,|\Delta|)$, so that the members at $x=\pm\delta$ relax much more slowly than $\bar H_S$ unless $g^2\gtrsim\delta$, as can be seen in Fig.~\ref{fig:dark_comparison}(b). The mechanism is clearest for $g^2\gg|\Delta|$. The dark mode then relaxes only by tunneling, with amplitude $\Delta/2$, into the bright mode, which the reservoir broadens over an energy $g^2$. The common shift moves both modes together, whereas $n_-$ detunes the dark level by $x$ from the bright one. Expanding Eq.~\eqref{eq:app_dark_eigs} gives the slow rate ${\rm Re}\,\lambda_+\simeq-\Delta^2g^2/[4\hbar(g^4+x^2)]$, a Lorentzian of width $g^2$ in the detuning, which reduces to the rate $\Delta^2/(4\hbar g^2)$ of Sec.~\ref{subsec:dqd} at $x=0$. Members detuned by $|x|\gtrsim g^2$ are thus off resonance and relax more slowly, so that the whole family relaxes at the rate of $\bar H_S$ only when $g^2\gtrsim\delta$. At strong coupling, the targets of both families depend only on the dark-mode energy, which changes by $x/2$ along the common shift and by $x$ along $n_-$. The bound along $n_-$ therefore approaches twice its common-shift value, as for the locally exact bound.

\subsection{Benchmark with speed limits} At finite $\Lambda$, the reservoir acts on the dots only through the collective mode $c=\Lambda_0^{-1}\sum_kt_kb_k$, with $\Lambda_0^2=\sum_k|t_k|^2=\Lambda/2$, so that $V=g\Lambda_0\sum_n(d_n^\dagger c+c^\dagger d_n)$. We solve the exact dynamics of the dots and $c$, with all remaining reservoir modes entering through the thermal noise of the Lorentzian continuum; no reservoir discretization is involved. The covariance of the dots and $c$ also yields $\delta E_S$ exactly, because $H_S+V$ acts only on these modes.

\paragraph*{Scaling of the speed limits.} In the factorized initial state, the dots and $c$ are uncorrelated, so that $\langle V\rangle=0$ and $H_S$ and $V$ are uncorrelated. Wick's theorem then gives
\begin{equation}
    \delta E_S^2(0)=\operatorname{Var}_{\rho_S(0)}(H_S)+g^2\Lambda_0^2\left[(1-n_c)\,\bm u^{\rm T}C(0)\,\bm u+n_c\big(N-\bm u^{\rm T}C(0)\,\bm u\big)\right],
    \label{eq:app_energy_variance}
\end{equation}
with $n_c=\langle c^\dagger c\rangle=\Lambda_0^{-2}\int\frac{d\Omega}{2\pi} \mathcal J(\Omega)f_\beta(\Omega)$. Since $\mathcal J$ is even and $f_\beta(\Omega)+f_\beta(-\Omega)=1$, $n_c=1/2$, and the interaction term equals $\operatorname{Var}(V)=Ng^2\Lambda/4$ for every preparation. The seminorm $\|H_S+V\|$ is the sum of the absolute eigenvalues of the one-particle matrix
\begin{equation}
    \mathcal{K}=\begin{pmatrix} h & g\Lambda_0\,\bm u\\ g\Lambda_0\,\bm u^{\rm T} & 0\end{pmatrix}.
    \label{eq:app_one_particle_matrix}
\end{equation}
For $g\Lambda_0\gg|\epsilon_n|$, the bright combination $\bm u/\sqrt N$ hybridizes with $c$ into two eigenvalues $\simeq\pm g\Lambda_0\sqrt N$, whereas the remaining $N-1$ eigenvalues stay of order $|\epsilon_n|$. Hence $\|H_S+V\|=g\sqrt{2N\Lambda}+O(|\epsilon_n|)$. Once $Ng^2\Lambda\gg\|H_S\|^2$ and $\Lambda\gg g^2$, the interaction dominates $\delta E_S$ along the whole evolution, and Eq.~\eqref{eq:ss_chain} gives, to leading order,
\begin{equation}
    \tau_{\rm SS}\simeq\max_{|x|\leq\delta}\frac{2\hbar\,\mathcal D_x}{g\sqrt{N\Lambda}}.
    \label{eq:app_ss_scaling}
\end{equation}
The static form, with $\|H_S+V\| \simeq g\sqrt{2N\Lambda}$, is smaller by a factor $\sqrt{2}$, so that both forms
decrease as $(g^2\Lambda)^{-1/2}$, whereas the MFG targets, and hence $\mathcal D_x$ and $\chi_\delta$, converge as $\Lambda\to\infty$.

\begin{figure}[t]
    \centering
    \includegraphics[width=\textwidth]{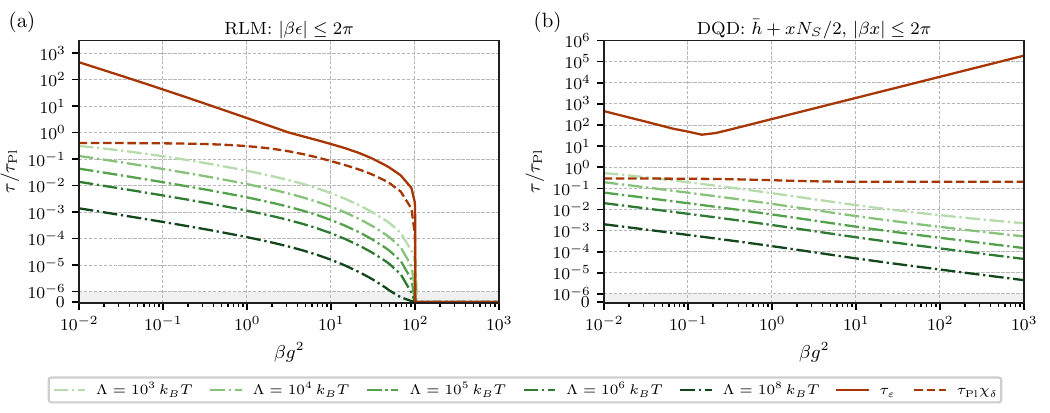}
    \caption{\textbf{Sensitivity of the speed limit to the bandwidth.}
    (a)~RLM and (b)~DQD families of Fig.~\ref{fig:finite_error} at accuracy $\varepsilon=0.05\,\varepsilon_{\max}$: Shiraishi--Saito bound $\tau_{\rm SS}$ of Eq.~\eqref{eq:ss_bound} (dash-dotted) in the Lorentzian model of width $\Lambda$, thermalization time $\tau_\varepsilon$ (solid), and bound $\tau_{\rm Pl}\chi_\delta$ (dashed). Within the wide-band regime, $\tau_\varepsilon$ and $\chi_\delta$ do not depend on $\Lambda$ on the scale of the figure; the curves shown are for $\Lambda=10^8k_BT$. For $\Lambda\leq10^4k_BT$, the strongest couplings shown violate $\Lambda\gg g^2$. The vertical axes are linear below $10^{-6}$, so that exact zeros remain visible.}
    \label{fig:speedlimit_sensitivity}
\end{figure}

\paragraph*{Comparison along the coupling.} Figure~\ref{fig:speedlimit_sensitivity} shows $\tau_{\rm SS}$ for the families of Fig.~\ref{fig:finite_error} at several bandwidths. In both models it decreases by an order of magnitude for every hundredfold increase of $\Lambda$, as predicted by Eq.~\eqref{eq:app_ss_scaling}, whereas $\tau_\varepsilon$ and $\chi_\delta$ remain unchanged.

In the RLM, $\tau_{\rm SS}$ lies below $\tau_{\rm Pl}\chi_\delta$ throughout the wide-band regime. At the threshold coupling $g_c^2$ of Sec.~\ref{subsec:rlm}, both bounds vanish together with $\tau_\varepsilon$. Since the preparation is the geodesic midpoint of the extreme targets, $\max_xD(\rho_S(0),\tilde\omega_x)=D(\tilde\omega_{-\delta},\tilde\omega_\delta)/2$, so that $\mathcal D_x=0$ for all members exactly when $\chi_\delta=0$. Both bounds then correctly signal that the task has become trivial.

The DQD shows where the speed limit ceases to be informative. As the coupling increases, $\tau_\varepsilon$ grows because the dark mode relaxes on the time scale $2\hbar g^2/\Delta^2$ (Sec.~\ref{subsec:dqd}), and $\chi_\delta$ remains finite because the dark mode retains the Hamiltonian information. The speed limit instead decreases as $(g^2\Lambda)^{-1/2}$ and vanishes at strong coupling, precisely where thermalization is slowest. It measures how fast the reduced state could move under $H_S+V$, whose fluctuations come from the bright mode and the reservoir, and it is blind to the dark mode, which controls both the relaxation and the equilibrium information. Our bound does not capture the slowing down either, since it stays far below $\tau_\varepsilon$, but it does not vanish.

The speed limit is not always weaker. When the fluctuations of $V$ do not dominate those of $H_S$, i.e., at weak coupling and in a narrow band, it can exceed $\tau_\mathrm{Pl}\chi_\delta$. In the DQD, this happens for $\Lambda=10^3k_BT$ at the weakest couplings of Fig.~\ref{fig:speedlimit_sensitivity}(b), and in narrower bands outside the wide-band regime. Both bounds hold for the same $\tau_\varepsilon$, and the larger of them applies.

\end{document}